\documentclass[fleqn,usenatbib]{mnras}

\usepackage{newtxtext,newtxmath}

\usepackage[T1]{fontenc}
\usepackage{ae,aecompl}

\usepackage{graphicx}	
\usepackage{amsmath}	
\usepackage{pdflscape}  
\usepackage{threeparttable}
\usepackage{afterpage}

\graphicspath{{Images/}}

\title[A Systematic Survey for z $<$ 0.04 Changing-Look AGNs]{A Systematic Survey for z $<$ 0.04 Changing-Look AGNs}

\author[M. R. Senarath et al.]{
Madhooshi R. Senarath,$^{1}$\thanks{madhooshi.senarath@monash.edu}
Michael~J.~I. Brown,$^{1}$
Michelle E. Cluver,$^{2,3}$
\newauthor
Thomas H. Jarrett,$^{5}$
Christian Wolf,$^{4}$
Nicholas P. Ross,$^{6}$ 
John R. Lucey,$^{7}$
\newauthor
Vaishali Parkash,$^{1}$
Wei J. Hon,$^{8}$
\\
$^{1}$School of Physics and Astronomy, Monash University, Clayton, Victoria 3800, Australia\\
$^{2}$Centre for Astrophysics and Supercomputing, Swinburne University of Technology, Hawthorn, Victoria, 3122, Australia\\
$^{3}$Department of Physics and Astronomy, University of the Western Cape, Robert Sobukwe Road, Bellville, 7535, South Africa\\
$^{4}$Research School of Astronomy and Astrophysics, Australian National University, Canberra, ACT 2611, Australia\\
$^{5}$Department of Astronomy, University of Cape Town, Private Bag X3, Rondebosch, 7701, South Africa\\
$^{6}$Institute for Astronomy, University of Edinburgh, Royal Observatory, Blackford Hill, Edinburgh EH9 3HJ, United Kingdom\\
$^{7}$ Centre for Extragalactic Astronomy, University of Durham, Durham DH1 
3LE, United Kingdom\\
$^{8}$ School of Physics, University of Melbourne, Parkville, Victoria 3010, Australia \\
}

\date{Accepted XXX. Received YYY; in original form ZZZ}

\pubyear{2020}

\begin{document}
\label{firstpage}
\pagerange{\pageref{firstpage}--\pageref{lastpage}}
\maketitle

\begin{abstract}

We have conducted a systematic survey for {\it z} $<$ 0.04 active Galactic nuclei (AGNs) that may have changed spectral class over the past decade. We use SkyMapper, Pan-STARRS and the {V{\'e}ron-Cetty} \& {V{\'e}ron} (2010) catalogue to search the entire sky for these ``changing-look'' AGNs using a variety of selection methods, where Pan-STARRS has a coverage of 3$\pi$ steradians (sky north of Declination $-30^\circ$) and SkyMapper has coverage of $\sim$ 21,000$~\rm{deg^2}$ (sky south of Declination $0^\circ$). We use small aperture photometry to measure how colour and flux have changed over time, where a change may indicate a change in spectral type. Optical colour and flux are used as a proxy for changing H$\alpha$ equivalent width, while \textit{WISE} 3.4 $\mu$m flux is used to look for changes in the hot dust component. We have identified four AGNs with varying spectra selected using our optical colour selection method. Three AGNs were confirmed from recent observations with  WiFeS on the 2.3~m telescope at Siding Spring and the other was identified from archival spectra alone. From this, we identify two new changing look AGNs; NGC 1346 and 2MASX J20075129-1108346. We also recover Mrk 915 and Mrk 609, which are known to have varying spectra in the literature, but they do not meet our specific criteria for changing look AGNs.

\end{abstract}

\begin{keywords}
galaxies: active -- galaxies: Seyfert -- methods: Observational
\end{keywords}




\section{Introduction}


The classic dichotomy of Active Galactic Nuclei (AGNs) classifies their optical spectra as having either broad or narrow emission lines, type 1 and type 2, respectively, with some intermediate classes containing both emission line components \citep{Sey1943,Wee1976,Ost1977,Ost1981}. The widely used unified model of AGNs proposes that observed AGN type/classes are a single type of object, observed at different orientations along the line of sight \citep{Ost1989,Ant1993}. We can directly observe both the broad line region (BLR) and narrow line region (NLR) in type 1 Seyferts. Whereas, in type 2 Seyferts, the light from the broad line region is absorbed by the dusty torus and is not visible in the optical (although it is observable in the IR), while the light from the narrow line region is scattered. Intermediate type 1 Seyferts can have both narrow and broad emission lines, type 1.5 Seyferts have narrow lines with obvious broad H$\alpha$ and H$\beta$ components, type 1.8s have narrow lines with a broad H$\alpha$ component and a recognisable broad H$\beta$ component and type 1.9s contain narrow lines with only H$\alpha$ line being broad \citep{Ost1977,Ost1981}.


Different wavelengths probe different regions of an AGN. The infrared (IR) wavelength range is sensitive to thermal emission from warm dust, which is often attributed to the torus that can obscure the ultraviolet and optical emission \citep[e.g.,][]{Pad2017}. The optical and ultraviolet (UV) bands probe emission from the accretion disk and fast moving gas (1000 - 10,000 km sec$^{-1}$) in the BLR, but the UV and optical emission from these regions can be obscured by dust. The X-ray band traces the emission of the hot corona and the ionized reflection of the X-ray continuum from distant neutral material like the molecular torus, the BLR and NLR or the accretion disk \citep{Ant1993, Geo1991,Jaf2004,Mei2007,Bia2008}. X-rays from AGNs are believed to be a result of inverse Compton scattering of the photons in the accretion disk by the hot corona. 


Changing-look AGNs (CLAGNs) are Seyferts and quasars where the spectral type changes from broadline to narrow line and vice versa. Given that the size of the torus is on the order of 1~pc and the relevant velocities are ${\rm< 10^4 ~km~ s^{-1}}$, one might expect CLAGNs to take ${\sim 10^3}$ years to change spectral class in the optical. We may expect variability on the viscous timescale, which for AGNs is on the order of ${\sim 10^3 - 10^5}$ years \citep{Sie1996}. However \citet{Toh1976}, \citet{Pen1984}, \citet{Tra1992}, \citet{Sto1993}, \citet{Era2001}, \citet{Mar2012}, \citet{Mar2013}, \citet{Den2014}, \citet{Sha2014}, \citet{Guo22016}, \citet{Okn2017}, \citet{Ros2018}, \citet{Nod2018} and \citet{Hon2020} for example, have identified CLAGNs that change spectral type in only a few years. CLAGNs may therefore be more common than previously believed.  

Examples of low redshift CLAGNs from the literature include Mrk 833 \citep{Can2018}, which changed from type 1.9 to type 1.8, NGC 7603 \citep{Toh1976}, which changed from type 1 to type 1.5, Mrk 372 \citep{Pen1984}, which changed from type 1.5 to type 1.9 and NGC 1566 \citep{Okn2018}, which changed from type 1.9 to type 1.2. While there isn't a strict definition in the literature, for consistency with the literature we classify objects as CLAGN if the broadline components completely disappears, a new broad line component appears and/or if the \citet{Ost1977} and \citet{Ost1981} spectral type changes by more than 0.1 (that is a change from type 1.8 to 1.9 and 1.9 to 2.0 and vice versa is not significant enough to be classified as a CLAGN). Some objects in our sample do show interesting spectral variability while falling below our CLAGN thresholds, and we retain them in this work while not classifying them as CLAGNs. AGNs such as Mrk 883, which change type by only 0.1 do not meet our CLAGN criteria, however it is considered an CLAGN by \citet{Can2018}.

\subsection{Why CLAGN change spectral type}

There are two main reasons why AGNs change their spectral type. One scenario is an obscuring cloud crosses the line of sight, causing changes in observed light curves. \citet{Goo1989} and \citet{Guo22016} have found AGNs which vary due to obscuring clouds. In this case, light from the inner disk and BLR is obscured by the dusty cloud, causing broad lines to disappear from the spectra. IR emission is sensitive to thermal emission from warm dust, which is often attributed to the torus, as the change in spectral type in this scenario is caused by obscuration of the BLR, we do not expect to measure a change in IR. Thus, the spectra of the AGNs should return to their original state after a period of time. A likely example of such a CLAGN is quasar SDSS J231742.60+000535.1 \citep{Guo22016}, where the change in its spectral type was caused by rapid outflow or inflow with an obscuring cloud passing along the line of sight. This scenario is in agreement with the unified model as changes in the spectral type are due to changes along the line of sight.

The second and more complex cause for change in spectral type is due to changes in accretion rate of the central black hole, changes in accretion disk structure, or tidal disruptions \citep[]{Dex2011,Kel2011,Mer2015,Kok2015,Mac2016}. \citet{Ros2018} use models of the innermost stable circular orbit around a black hole to determine if this is a possible driver for changes in the spectra of SDSS J1100-0053, where the different models have combinations of zero torque, non-zero torque, spectral hardening factor\footnote{The spectral hardening factor, also referred to as the color correction, is used to interpret multi-temperature black body fitting results \citep{Dav2019}. Where for a canonical blackbody spectral, the spectral hardening factor is 1.} and radii. \citet{Ros2018} attributed the change in spectral type to mass flow rate switching from cold, high  mass flow rate to hot, low mass flow rate. Unlike the previous scenario, the change in spectral type is caused by changes in accretion. Thus, it does not agree with the unified model as the spectral type change is not caused by changes along the line of sight. 

\subsection{Known CLAGNs}

\begin{figure*}
\begin{center}
\includegraphics[width=1\textwidth,angle=0]{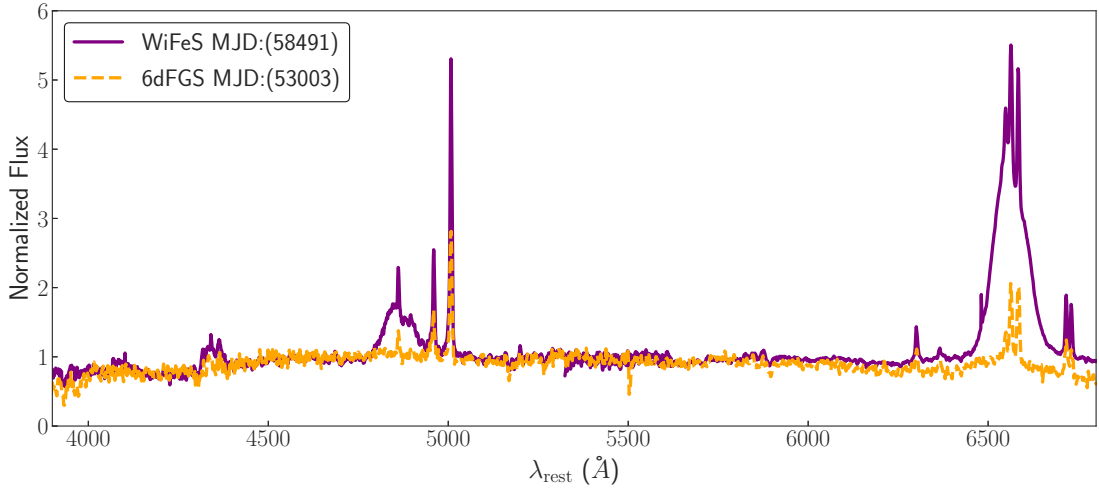}
\caption{Spectra of NGC 2617, a previously identified CLAGN with still evolving spectra from type 1.8 to type 1 Seyfert \citep{Okn2017}. \citet{Ver2010} identify the 6dFGS spectrum as type 1.8, while our WiFeS 2019 January shows NGC 2617 is currently a type 1, which agrees with observations from \citet{Okn2017}.
\label{fig:NGC2617}}
\end{center}
\end{figure*}

Until recently, the most common method by which studies have identified CLAGNs is by serendipity. For example, NGC 2617 is a CLAGN that was identified by \citet{Sha2014} after an outburst triggered a transient source alert, and the corresponding changing optical spectra are displayed in Figure~\ref{fig:NGC2617}. NGC 2992 was identified by \citet{Gil2000} using \textit{Beppo}SAX observations \citep{Sca1997} which caught a rise in nuclear emission from the AGN, and there was a corresponding change in the optical classification from type 1.9 to type 2. 

It is only recently that targeted searches for CLAGNs such as \citet{Lam2015}, \citet{Mac2016}, \citet{Rua2016}, \citet{Run2016}, \citet{Gez2017}, \citet{Yan2018}, \citet{Ste2018} and \citet{Mac2019} have been conducted. This is because there is now more readily available archival data and multi-epoch photometry such as NEOWISE \citep{Mai2014}, Pan-STARRS \citep{Cha2016}, SDSS \citep{Eis2011}, SkyMapper \citep{Wol2018} and GAIA \citep{Pru2016}. These targeted searches are focused mainly on detecting changing-look quasars that are well beyond {\it z} $\sim$ 0. That said, a number of changing-look Seyferts have been identified in the {\it z}~$<$~0.04 Universe, and these and their key details including previous and current type are summarised in Table~\ref{tbl:known}. The CLAGNs in Table~\ref{tbl:known} were identified by serendipity or X-ray monitoring of known AGNs.

\begin{table*}
\caption{Known {\it z}~$<$~0.04 CLAGNs and their respective types and references.}
\label{tbl:known}
 \scriptsize
\begin{tabular}{ccccccc}
\hline
ID & Ra(J2000) & Dec(J2000) & redshift & Previous type(s) & Current type& Reference\\ [0.5ex] 
    \hline
NGC~7603& 23h18m56.65s & +00d14m37.9s & 0.030 & 1 &1.5&\citet{Toh1976}\\
NGC~4151 & 12h10m32.65s & +39d24m20.7s & 0.003& 1, 2 &1.5&\citet{Pen1984}\\
NGC 2622 & 08h38m10.943s & +24d53m43.02s & 0.029& 1.8 & 1 & \citet{Goo1989}\\ 
Mrk 372 & 03h02m13.18s & -23d35m19.8s & 0.035 & 1.5 & 1.9 &\citet{Gre1991}\\
Mrk~993 & 01h25m31.47s & +32d08m10.5s & 0.016 & 1 & 1.9 &\citet{Tra1992}\\ 
NGC~1097 & 02h46m19.05s & -30d16m29.6s & 0.004 & 2 & 1 &\citet{Sto1993}\\
NGC~3065 & 10h01m55.30s & +72d10m13.0s & 0.007&2&1.2$^a$&\citet{Era2001}\\ 
NGC~2992 & 09h45m42.04s & -14d19m34.8s &0.008&1.9, 1.5&2&\citet{Gil2000} $\&$ \citet{Tri2008}\\
NGC~454E &  01h14m22.50s & -55d23m55.0s & 0.012&&2$^b$&\citet{Mar2012}\\ 
NGC 1365 & 03h33m36.45s & -36d08m26.3s & 0.005 &   &1.8$^b$& \citet{Mar2013} \\ 
Mrk~590 & 02h14m33.57s & -00d46m00.2s & 0.026 & 1 & 1.9-2 &\citet{Den2014}\\
Mrk 6 & 06h52m12.251s & +74d25m37.46s & 0.019& 2 & 1.5 & \citet{Kha1971}, \citet{Kha2011} $\&$ \citet{Afa2014}\\
ESO~362-G018 & 05h19m35.80s & -32d39m27.0s &0.012&1.5&2&\citet{Agi2017}\\
NGC~7582 & 23h18m23.62s & -42d22m14.0s &0.005&1&2&\citet{Bra2017}\\

NGC~2617 & 08h35m38.77s & -04d05m17.2s &0.014&1.8&1&\citet{Okn2017}\\
NGC~1566 &  04h20m00.41s & -54d56m16.1s &0.005&1.9&1.2&\citet{Okn2018}\\
Mrk~883 & 16h29m52.84s & +24d26m37.4s &0.037&1.9&1.8&\citet{Can2018}\\
HE 1136-2304 & 11h38m51.00s & -23d21m32.0d & 0.027 & 2 & 1.5 & \citet{Zel2018}\\
NGC 3516 & 11h06m47.490s & +72d34m06.88s & 0.009& 1 & 2 & \citet{Sha2019}\\
1ES 1927+654 & 19h27m19.54s & +65d33m54.2s&0.017&2&1&\citet{Tra2019}\\ 
NGC 1346 &03h30m13.27s & -05d32m36.3s 
& 0.014 & 1.8& 2 &\citet{me} $\&$ This work\\
2MASX J20075129-1108346 & 20h07m51.29s & -11d08m34.6s & 0.030 & 2 & 1.8 & This work\\

\hline
\end{tabular}
\begin{tablenotes}
\item[a]$^a$ While \citet{Era2001} don't explicitly state the spectral types of the changes, they do state that NGC 3065 went from lacking broad Balmer lines to containing broad Balmer lines.
\item[b] $^b$ NGC 1365 is an X-ray CLAGN. X-ray CLAGNs are characterised by rapid transitions between Compton-thick to Compton-thin, where this transition can be due to absorption by gas clouds passing along the line of sight or relativistic reflection on to the accretion disk \citep{Mar2013}. In the case of NGC 1365, the CLAGN classification is due to the reflection-dominated scenario.

\end{tablenotes}
\end{table*}

Motivated by the frequency of ad hoc nearby CLAGN identifications, we have conducted a systematic search for {\it z} $<$ 0.04 CLAGNs. Our redshift limit is chosen to keep H$\alpha$ within the $r$ band and increases availability of archival photometry and spectroscopy. We use the comprehensive \citet{Ver2010} catalogue to identify {\it z} $<$ 0.04 AGNs using SkyMapper \citep{Wol2018}, the Panoramic Survey Telescope and Rapid Response System \citep[Pan-STARRS,][]{Cha2016,Fle2016}, the Wide-field Infrared Survey Explorer \citep[\textit{WISE},][]{Wri2010} and Sloan Digital Sky Survey \citep[\textit{SDSS},][]{Aba2004}. We select CLAGN candidates meeting our colour and flux criteria, and then follow-up these candidates with archival and new spectroscopy. In Section~\ref{sec:criteria} we discuss the methods by which we selected CLAGN candidates and the effectiveness of each method.  In Section~\ref{sec:observations} we discuss what objects were observed. In Section~\ref{sec:results} we discuss the new CLAGNs we identified, including NGC 1346 which we also discuss in \citet{me}. We also discuss the possible reasons why these AGNs changed spectral type in Section~\ref{sec:results}.


\section{Candidate selection}\label{sec:criteria}

To select CLAGN candidates we use photometry from PanSTARRS, SkyMapper and the Wide-field Infrared Survey Explorer \citep[\textit{WISE},][]{Wri2010}. Our methods use optical and MIR fluxes and colours to search for variability in the optical continuum. Our first approach uses optical colours as a proxy of H$\alpha$ equivalent widths. Our second and third approaches use variability of optical and MIR fluxes to search for changes in accretion disk and hot dust component respectively. 

We use the MIR to identify changes in the dust near the accretion disk of the AGNs (presumably not associated with the larger torus). The optical and UV continuum probes emission from the disk while the optical and UV spectral lines probe ionised gas above the disk, and changes in the optical flux and color may indicate changes in spectral type resulting from changes in accretion or changing obscuration along the line of sight. We use optical colour and infrared fluxes and colours as proxies for H$\alpha$ emission and hot disk emission respectively, where variations would indicate the presence of a CLAGN. Therefore, we require photometry of known AGNs (the specifics of this are explained in Section~\ref{sec:thedata}). Following the photometric selection of CLAGN candidates we obtained follow-up spectroscopy.

\subsection{Imaging surveys and catalogues}\label{sec:thedata}

We select known {\it z} $<$ 0.04 AGNs from \citet{Ver2010}, an extensive compilation of AGNs (particularly at low redshift). We then measure optical and infrared photometry of these AGNs with SkyMapper, Pan-STARRS, and WISE. Possible CLAGNs are then selected with the colour and flux criteria described Sections~\ref{sec:opticalcolour},~\ref{sec:opticalflux} and \ref{sec:IRflux}. The SkyMapper survey (with passbands $uvgriz$) contains $\approx$ 280 million objects and has a coverage area of almost the entire Southern sky. Pan-STARRS on the other hand, surveys the sky north of Declination $-30^\circ$ (passbands $grizy$). Together the two surveys provide data for the entire sky. 

The depth of the optical catalogues are $r \sim 21.7$ mag for SkyMapper and $r \sim 23.2$ mag for Pan-STARRS, which is more than sufficient to detect all the z $<$ 0.04 \citet{Ver2010} AGNs. For our analysis we use small aperture photometry. For Pan-STARRS (PS1) we measure photometry in 3$^{\prime\prime}$ diameter apertures on stack images. For SkyMapper we use the DR1 5$^{\prime\prime}$ diameter aperture photometry. For SDSS we use the fibre magnitudes (3$^{\prime\prime}$ in diameter). To photometrically identify AGNs that may have a varying hot dust component, we use photometry drawn from the \textit{WISE} and NEOWISE surveys \citep{Mai2014}. NEOWISE measures photometry in the W1 and W2 bands and surveys the entire sky at a cadence of 6 months, and has been doing so since \textit{WISE} was brought out of hibernation in late 2013. We present a subset of our CLAGN candidate catalogues in Tables~\ref{tbl:candidates} and \ref{tbl:candidatespanstarrs} for SkyMapper and Pan-STARRS, respectively, which contain all the {\it z} $<$ 0.04 AGN with information on the selection criteria and whether or not each galaxy meets the criteria. They also contain archival spectra references for the Seyferts that meet the CLAGN candidate criteria.

\subsection{Optical colour selection}\label{sec:opticalcolour}
CLAGNs should change colour due to varying H$\alpha$ strength, thus we use $r$ - $i$ as a proxy for H$\alpha$ equivalent width. Our optical colour selection assumes that type 1 and type 2 AGNs have relatively blue and red $r$ - $i$ colours, respectively, resulting from the equivalent width of the H$\alpha$ emission line. As the Pan-STARRS and SkyMapper bands differ from each other, the $r$ - $i$ colours they measure for individual AGNs will differ. As a consequence we compare {\it z} $<$ 0.04 AGNs that appear in both catalogues, we find a linear relation as a function of $r$ - $i$ colour. We apply this relation when determining our $r$ - $i$ colour criteria for our Pan-STARRS and SkyMapper catalogues.
 
\begin{figure} 
    \centering
    \includegraphics[width=\columnwidth]{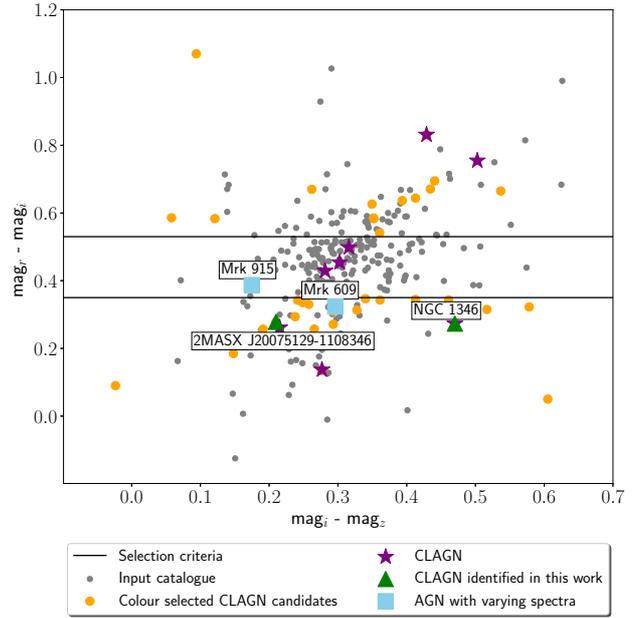}
    \caption{ Skymapper $riz$ colours for our CLAGN candidates with the other z $<$ 0.04 AGNs displayed in the background. We select \citet{Ver2010} type 1s with $r-i$ $>$ 0.53 mag and type 2s with $r$ - $i$ $<$ 0.35 mag as possible CLAGN candidates, these are indicated in orange.}
    \label{fig:Skycandidates}
\end{figure}

\begin{figure} 
    \centering
    \includegraphics[width=\columnwidth]{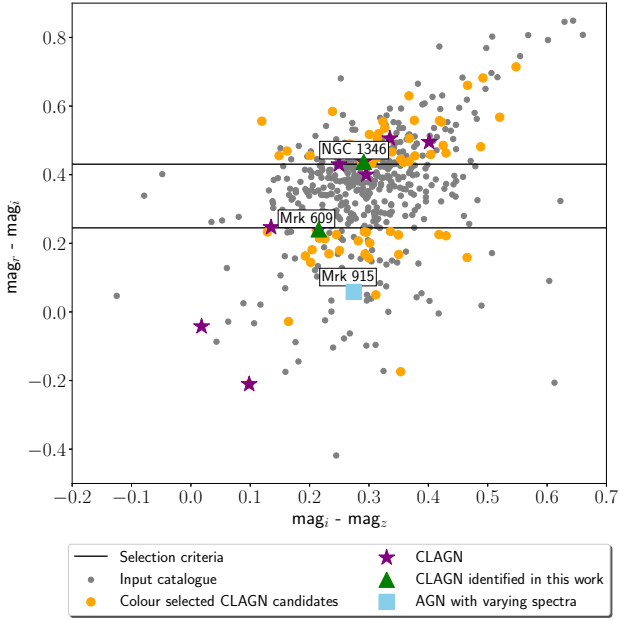}
    \caption{Pan-STARRS catalogue $riz$ colours for our CLAGN candidates with the other z $<$ 0.04 AGNs displayed in the background. We select \citet{Ver2010} type 1s with $r$ - $i$ $>$ 0.43 mag and \citet{Ver2010} type 2, type 1.8 and type 1.9 with $r$ - $i$ $<$ 0.25 mag as potential CLAGN candidates, these are indicated in orange.}
    \label{fig:Pancandidates}
\end{figure}

For our SkyMapper catalogue we select blue Seyfert type 2s with $r$ - $i$ $<$ 0.35 mag and red type 1s with $r$ - $i$ $>$ 0.53 mag (after removing flagged objects with spurious photometry). For our Pan-STARRS sample we select blue type 2s with $r$ - $i$ $<$ 0.25 mag and red type 1s with $r$ - $i$ $>$ 0.43 mag (after removing flagged objects with spurious photometry). This selects a total of 109 candidates where we select 40 and 69 AGNs in SkyMapper and Pan-STARRS, respectively. These candidates are displayed in Figure~\ref{fig:Skycandidates} and Figure~\ref{fig:Pancandidates}. 

We further refine this candidate list by inspecting the archival images and spectra, including identifying changes in the ${\rm H\alpha}$ and ${\rm H\beta}$ emission lines. We select candidates to observe on the basis that they have more than one archival spectra spectra and there is variation in emission line widths that appear to be changing in the last 10 years (after 2008), these archival spectra are referenced in Tables~\ref{tbl:candidates} and \ref{tbl:candidatespanstarrs}. It should be noted that 46$\%$ of the candidates selected using the mentioned colour criteria have no readily available archival spectra or have just one readily available spectrum. Also 26$\%$ of candidates did not have archival spectra taken in the last 10 years. AGNs that fall into these categories are not selected for observations. 

Of the 22 known {\it z} $<$ 0.04 changing-look Seyferts, 10 are known to have changed optical spectral class between 1998 and 2015, where 2015 is defined by the end of the SkyMapper observations for the data release we are using (DR1) \citep{Wol2018} and 1998 is defined by the beginning of the SDSS imaging we are using \citep{Eis2011}, while Pan-STARRS PS1 data was obtained between 2009 and 2014 \citep{Cha2019}. Our photometric selection has recovered five of these known CLAGNs; Mrk 883, Mrk 590, NGC 4151, NGC 7282 and NGC 2617 (further information on these CLAGNs appear in Table~\ref{tbl:known}).

\subsection{Optical flux variability selection}\label{sec:opticalflux}

As with $r$ - $i$ colour selection, we utilise the r-band flux variability to detect the changing H$\alpha$ emission of CLAGNs. As, by definition this requires multiple $r$ band epochs, we have measured the variability of the \citet{Ver2010} z $<$ 0.04 AGNs in the $\approx$ 14,055 square degrees that have both SDSS and Pan-STARRS photometry. It should also be noted that the time between SDSS and Pan-STARRS observations can be as little as a few years or longer than a decade, while changes in Seyfert spectra can take decades to occur, or in some cases can be limited to just a few years \citep[i.e, tidal disruptions,][]{Guo22016}.

\begin{figure}
    \centering
    \includegraphics[width=\columnwidth]{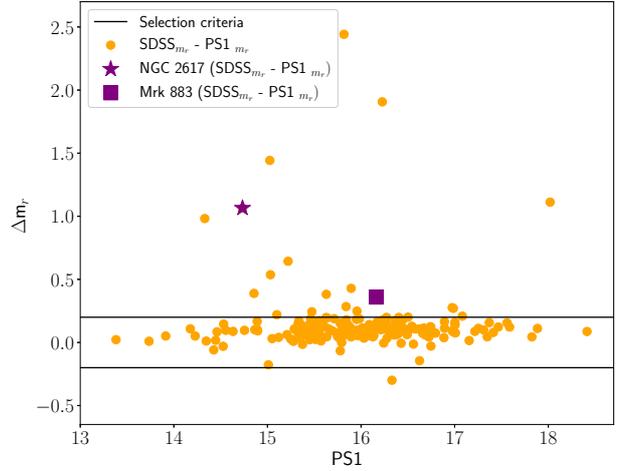}
    \caption{The $r$-band magnitude as a function of apparent magnitude for \citet{Ver2010} type 1.8s, type 1.9s and type 2s, measured with SDSS and Pan-STARRS 3$^{\prime\prime}$ aperture photometry. Known CLAGNs NGC 2617 and Mrk 883 are highlighted, NGC 2617 became ~1 magnitude brighter in $r$-band between the SDSS and Pan-STARRS imaging surveys. }
    \label{fig:PanS2s}
\end{figure}

We found that type 1.8s, 1.9s and type 2s typically showed variability $\Delta$m$_r$ $<$ 0.2 mag, and these are plotted in Figure~\ref{fig:PanS2s}. However type 1s showed a greater variability making it impractical to select type 1s fading into type 2 on the basis of $r$ band flux variability alone. Of the 335 type 1.8s, 1.9s and type 2s with both SDSS and Pan-STARRS photometry, we identified 22 potential CLAGNs using $\Delta$m$_r$ $>$ 0.2 mag. It should be noted that as these objects are relatively bright, the photon shot noise is relatively small and the scatter is dominated by PSF differences, zeropoint errors, filter curve differences and AGN variability. The dominant source of error is systematic errors rather than easily quantifiable errors, and thus we have not included individual error bars into Figure~\ref{fig:PanS2s}.

After further investigation of the 22 potential CLAGNs, four CLAGN candidates were removed as their measurement of variability resulted from centroid errors, where two of these AGNs were in galaxy pairs. We also recover the known CLAGNs NGC 2617 \citep{Okn2017} and Mrk 883. Further inspection into archival spectra of the AGNs with $\Delta$m$_r$ $>$ 0.2 (NGC1048A, SDSSJ03205-0020, Zw497.016, NPM1G-16.0109, MCG+08.15.009, NGC4565, Mrk 1392, Mrk 673, Mrk 609, UGC 4145, Zw098.038, HS1656+3927, NGC6264, Mrk 248 and IC1725) showed all had archival spectra. Inspecting these archival spectra, we found no evidence of variation. As such, we did not undertake spectroscopic follow up of these AGNs.

\subsection{Infrared flux variability selection}\label{sec:IRflux}

\begin{figure}
    \centering
    \includegraphics[width=\columnwidth]{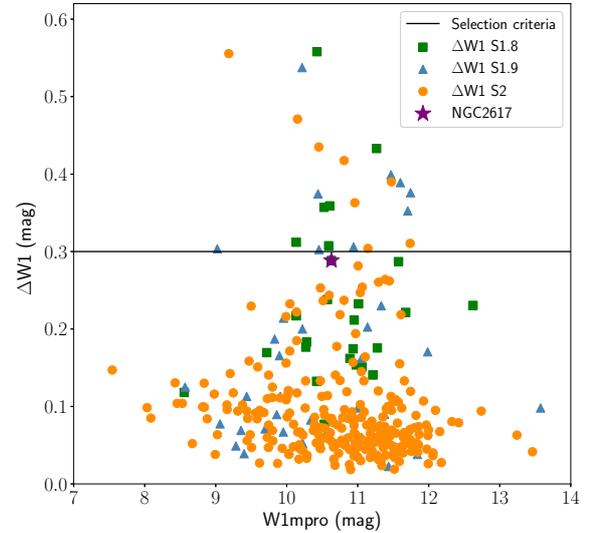}
    \caption{Change in NEOWISE W1 photometry as a function of W1 photometry. As type 1s vary at $\sim 3.5~{\rm \mu m}$  without changing spectral type, we can only plot and draw CLAGN candidates from AGNs classes type 1.8s, 1.9s and type 2.0s. The purple star is NGC 2617, a known CLAGN. }
    \label{fig:Neowise}
\end{figure}

We use NEOWISE variability to search for {\it z} $<$ 0.04, \citet{Ver2010} AGNs where the contribution of hot dust to the SEDs may be changing. Unlike optical flux variability selection where we measure the difference between SDSS and Pan-STARRS photometry, in this scenario we measure the change in magnitude over time using different epochs of NEOWISE. Each NEOWISE epoch has multiple photometry measurements, so we use the median photometry measurement for each epoch and measure the difference between the highest and lowest magnitude epochs, we present this in Figure~\ref{fig:Neowise}.

Most of the AGNs that display a high change are type 1 Seyferts which naturally vary over time.
Variability on the timescale of months or a few years indicated some emission is occurring close to the central engine \citep{Rui1986,Bur2015}, and the varying MIR emission is attributed to a dusty wind in the AGNs polar region \citep{Hon2013}. As with the previous CLAGN candidate selection methods, we inspect the archival spectra of AGNs where 1.8s, 1.9s and type 2s had $\Delta$W1 $>$ 0.3 mag. In all instances where we identified a type 2 with \textit{WISE} variability and spectra more recent than 2017, we found that the recent spectra still exhibited narrow lines (i.e., NGC 4135, NGC 6230, IC 1495 and Mrk 670). Thus, we did not undertake any spectroscopic follow up of AGNs on the basis of infrared variability alone. 

\afterpage{
\begin{landscape} 
\begin{table}
\caption{Sample taken from our SkyMapper catalogue used to select CLAGN candidates. Note that the full table is available online.}
\label{tbl:candidates}
\centering
\scriptsize
\begin{tabular}{|*{19}{c|}}
\hline
\multicolumn{5}{|c}{} & \multicolumn{3}{|c}{SkyMapper} & \multicolumn{4}{|c}{SDSS} & 
\multicolumn{4}{|c}{Pan-STARRS$^a$} & \multicolumn{3}{|c}{}  \\ \hline 
RA & Dec &Name & z & VCV & r & i & z & MJD & r & i & z & MJD& r & i & z & Flag$^b$ & Spectra &Spec.\\ 
(deg) &(deg) &&& Spec. Type& (mag) & (mag) & (mag) & & (mag) & (mag) & (mag) & & (deg) & (deg) &  (deg)&  &Notes$^c$& Sources$^d$\\\hline
0.8839&-10.7446&NGC7808&0.03&1&15.11&14.62&14.39&51792&15.92&15.48&15.13&55866&15.85&15.38&15.09&0&3&2,3\\
2.7260&-21.0675&ESO538-G25&0.03&2&15.77&15.32&14.93&53353&16.67&16.20&15.78&56028&16.56&16.08&15.78&0&2&2\\
2.7776&-12.1079&MARK938&0.02&2&14.48&13.95&13.69&0&0.00&0.00&0.00&56170&15.44&15.03&14.56&0&2&2\\
6.8176&-1.7802&NGC118&0.04&2&14.52&14.16&13.90&54769&15.29&14.90&14.65&55618&15.26&14.89&14.69&0&0&0\\
8.5568&-21.4386&ESO540-G01&0.03&1.8&14.61&14.32&14.08&53995&15.38&15.07&14.79&55913&15.33&15.03&14.84&2&0&0\\
8.9533&-13.6106&NGC166&0.02&2&15.43&15.04&14.55&0&0.00&0.00&0.00&56123&16.34&15.94&15.54&0&0&0\\
9.3992&0.2807&MARK955&0.04&2&15.14&14.72&14.46&52231&15.94&15.56&15.26&55541&15.87&15.43&15.22&0&2&3\\
10.7200&-23.5410&NGC235&0.02&1&14.50&14.04&13.66&53995&15.07&14.73&14.42&55937&14.99&14.65&14.35&0&2&5\\
11.0909&-17.3512&ESO540-G17&0.03&2&16.52&16.10&15.73&0&0.00&0.00&0.00&56026&17.41&16.99&16.66&0&2&2\\
13.3747&-8.7677&NGC291&0.02&2&15.55&14.88&14.74&51814&16.38&15.98&15.57&55768&16.32&15.96&15.59&0&3&2,3\\
13.7271&-32.0317&ESO411-G029&0.03&2&15.60&15.01&14.75&0&0.00&0.00&0.00&0&0.00&0.00&0.00&0&2&2\\
14.5930&-36.6601&ESO351-G025&0.04&2&16.20&15.77&15.45&0&0.00&0.00&0.00&0&0.00&0.00&0.00&0&2&2\\
16.3203&-58.4375&ESO113-G10&0.03&1.8&14.89&14.53&14.29&0&0.00&0.00&0.00&0&0.00&0.00&0.00&0&0&0\\
18.0802&-32.0612&NGC427&0.03&1.2&15.52&15.22&14.99&0&0.00&0.00&0.00&0&0.00&0.00&0.00&0&2&2\\
18.2025&-0.2902&SDSSJ01128-0017&0.02&2&14.44&13.95&13.63&52963&15.16&14.74&14.35&55635&15.11&14.65&14.42&0&2&3\\
18.5293&-32.6509&IC1657&0.01&2&15.39&14.68&14.55&0&0.00&0.00&0.00&0&0.00&0.00&0.00&0&3&2,4\\
18.7029&-0.4961&UGC793&0.03&1.5&16.09&15.59&15.30&53272&16.91&16.54&16.23&55834&16.81&16.46&16.27&0&3&3,32\\
18.9802&-50.1894&ESO195-G35&0.02&2&14.84&14.43&14.16&0&0.00&0.00&0.00&0&0.00&0.00&0.00&0&2&2\\
20.0821&-44.1287&ESO244-G17&0.02&1.5&15.47&15.13&14.84&0&0.00&0.00&0.00&0&0.00&0.00&0.00&0&0&0\\
20.8383&-1.9766&UM319&0.02&2&16.19&15.73&15.43&54770&17.08&16.74&16.35&55913&17.03&16.65&16.38&0&2&2\\
20.9766&-35.0654&NGC526A&0.02&1.9&14.93&14.42&14.00&0&0.00&0.00&0.00&0&0.00&0.00&0.00&0&3&2,5\\
23.4906&-36.4932&NGC612&0.03&2&15.27&14.64&14.25&0&0.00&0.00&0.00&0&0.00&0.00&0.00&0&3&2,5\\
25.9065&-33.7054&ESO353-G38&0.03&2&15.11&14.64&14.40&0&0.00&0.00&0.00&0&0.00&0.00&0.00&0&0&0\\
25.9907&2.3499&MARK573&0.02&1&14.46&14.12&13.89&54742&15.12&14.95&14.62&56011&15.03&14.88&14.57&0&2&4\\
27.9245&-36.1878&ESO354-G04&0.03&1&15.13&14.64&14.37&0&0.00&0.00&0.00&0&0.00&0.00&0.00&0&2&2\\
28.2042&-3.4468&MCG-01.05.047&0.02&2&15.94&15.43&15.10&54770&16.92&16.44&16.02&56294&17.50&16.79&16.28&0&3&2,5\\
29.9634&-6.8404&IC184&0.02&2&15.55&15.15&14.78&54832&16.48&16.10&15.78&56178&16.41&15.96&15.74&0&3&2,15\\
30.2769&-6.8159&NGC788&0.01&1&14.39&13.99&13.70&54832&15.28&14.93&14.57&56178&15.19&14.78&14.57&0&2&5\\
32.3525&-10.1359&NGC835&0.01&2&13.84&13.37&13.07&51813&14.68&14.25&13.93&56091&14.50&14.12&13.85&0&3&2,4\\
33.4098&-0.7173&SDSSJ02136-0043&0.02&1&15.21&14.87&14.71&52963&16.17&15.79&15.45&56254&16.06&15.65&15.46&0&2&3\\
33.6399&-0.7667&MARK590&0.03&1&14.58&14.08&13.77&52963&15.33&14.94&14.60&56311&15.26&14.83&14.58&0&2&3\\
37.5230&-8.9982&MARK1044&0.02&1&14.49&14.37&14.08&54057&14.72&14.75&14.57&56170&14.71&14.59&14.34&0&3&2,5\\
37.9625&-36.6721&IC1816&0.02&2&14.81&14.42&14.14&0&0.00&0.00&0.00&0&0.00&0.00&0.00&0&3&2,4,5,20\\
38.2552&0.4208&UGC2024&0.02&2&14.96&14.59&14.30&0&0.00&0.00&0.00&56284&15.43&15.09&14.92&0&3&2,3\\
38.4651&1.1371&SDSSJ02338+0108&0.02&1&15.24&14.79&14.54&54715&15.92&15.52&15.17&56262&15.83&15.41&15.21&0&2&3\\
\hline
\end{tabular}
\begin{tablenotes}
\item[a]$^a$ NOTE: The Pan-STARRS photometry in this table has been measured by us using the Pan-STARRS cutouts (where available) with a 3$^{\prime\prime}$ diameter aperture for the AGNs in our Skymapper catalogue. 

\item[b]$^b$ Optical colour selection flag where 1 indicates possible narrowing spectra and 2 indicates possible broadening spectra. 

\item[c]$^c$ CLAGN spectra flag where: 0 No archival spectra online, 1 Varying in archival and/or WiFeS spectra, 2 Only one archival spectrum found and no WiFeS, 3 Two or more archival spectra, not varying in WiFeS and archival spectra, 4 No archival spectra found from the last 10 years, but shows signs of varying before then and 5 No archival spectra found from the past 10 years, does not show signs of varying before then.

\item[d]$^d$ If Spectra Notes is 0, then this is also 0. 
1 WiFeS
2 6dFGS
3 SDSS
4 S7
5 BASS
6 2dFGRS
7 \citet{Ho1995} 
8 MaNGA
9 \citet{Fos1982} 
10 \citet{Phi1983} 
11 \citet{Ken1984} 
12 \citet{Ber1986} 
13 \citet{Ver1986} 
14 \citet{Ver19862} 
15 \citet{Kol1987} 
16 \citet{Mai1987} 
17 \citet{Rud1988} 
18 \citet{Mor1988} 
19 \citet{Sto1990} 
20 \citet{Win1992} 
21 \citet{Gri1992} 
22 \citet{Mor1994} 
23 \citet{Cru1994} 
24 \citet{Goo1995} 
25 \citet{Mai1996} 
26 \citet{Mor1996} 
27 \citet{Rei1996} 
28 \citet{Sca1996} 
29 \citet{Coz1997} 
30 \citet{Pie1998} 
31 \citet{Fra2000} 
32 \citet{Jan2000} 
33 \citet{Kew20012} 
34 \citet{Reu2003} 
35 \citet{Mar2004} 
36 \citet{Geo2004} 
37 \citet{Mas2006} 
38 \citet{Mas20062} 
39 \citet{Mou2006} 
40 \citet{Ho2009} 
41 \citet{Tri2010} 
42 \citet{Dop2014} 
43 \citet{Sch2016} 
44 \citet{Ram2016} 
45 \citet{Tho2017} 
\end{tablenotes}
\end{table}
\end{landscape}
}

\afterpage{
\begin{landscape} 
\begin{table}
\caption{Sample taken from our Pan-STARRS (PS1) catalogue used to select CLAGN candidates. Note that the full table is available online.}
\label{tbl:candidatespanstarrs}
\centering
\scriptsize
\begin{tabular}{|*{15}{c|}}
\hline
\multicolumn{5}{|c}{} & \multicolumn{3}{|c}{Pan-STARRS (PS1)} & \multicolumn{4}{|c}{SDSS} & 
 \multicolumn{3}{|c}{}  \\ \hline 
RA & Dec &Name & z & VCV & r & i & z & MJD & r & i & z &  Flag$^a$ & Spectra &Spec.\\ 
(deg) &(deg) &&& Spec. Type& (mag) & (mag) & (mag) & & (mag) & (mag) & (mag) &   &Notes$^b$& Sources$^c$\\\hline
0.4933&36.6489&Zw517.014&0.032&2&16.13&15.78&15.53&0&0.00&0.00&0.00&0&0&0\\
0.6100&3.3517&MARK543&0.026&1.5&15.75&15.64&15.38&54742&15.76&15.58&15.31&0&0&0\\
0.7900&21.9600&MARK334&0.022&1.8&15.51&15.27&14.89&54849&15.54&15.27&14.86&0&0&0\\
0.8837&-10.7447&NGC7808&0.029&1&15.86&15.38&15.10&51792&15.92&15.48&15.13&1&3&2,3\\
1.5813&20.2031&MARK335&0.026&1&14.35&14.45&14.21&55119&14.43&14.90&14.45&0&2&5\\
2.7262&-21.0675&ESO538-G25&0.026&2&16.54&16.07&15.77&53353&16.67&16.20&15.78&0&2&2\\
2.7775&-12.1075&MARK938&0.019&2&15.23&14.80&14.32&0&0.00&0.00&0.00&0&2&2\\
4.5979&30.0631&NGC71&0.022&2&15.95&15.48&15.18&55122&16.02&15.59&15.21&0&0&0\\
6.8175&-1.7797&NGC118&0.037&2&15.28&14.90&14.71&54769&15.29&14.90&14.65&0&0&0\\
8.5575&-21.4389&ESO540-G01&0.027&1.8&15.33&15.03&14.84&53995&15.38&15.07&14.79&0&0&0\\
8.9533&-13.6103&NGC166&0.020&2&16.33&15.94&15.54&0&0.00&0.00&0.00&0&0&0\\
9.3992&0.2808&MARK955&0.035&2&15.89&15.47&15.25&52231&15.94&15.56&15.26&0&2&3\\
9.7392&48.3372&NGC185&0.000&2&17.34&16.99&16.76&0&0.00&0.00&0.00&0&2&7\\
10.6846&41.2694&M31&0.000&2&12.52&100.00&12.21&0&0.00&0.00&0.00&2&3&7,39,50\\
10.7200&-23.5411&NGC235&0.022&1&15.00&14.67&14.37&53995&15.07&14.73&14.42&0&2&5\\
11.0908&-17.3517&ESO540-G17&0.031&2&17.34&16.94&16.62&0&0.00&0.00&0.00&0&2&2\\
11.8308&14.7033&MARK1146&0.039&1&16.77&16.25&15.95&51878&16.51&16.12&15.82&1&2&3\\
12.1967&31.9569&MARK348&0.014&1&15.63&15.36&15.08&55121&15.78&15.52&15.13&0&2&5\\
12.8958&29.4011&UGC524&0.036&1&15.77&15.18&14.94&55122&15.75&15.36&15.05&1&0&0\\
13.3742&-8.7678&NGC291&0.019&2&16.31&15.95&15.60&51814&16.38&15.98&15.57&0&3&2,3\\
13.7608&-19.0047&ESO541-G001&0.021&2&16.20&15.83&15.57&0&0.00&0.00&0.00&0&0&0\\
14.9171&15.3308&UGC615&0.018&2&15.63&15.19&14.90&51464&15.68&15.20&14.94&0&2&3\\
14.9721&31.8269&MARK352&0.015&1&15.87&15.47&15.52&55121&15.76&15.69&15.48&0&2&5\\
17.1983&-15.8425&IC78&0.040&2&16.04&15.57&15.12&0&0.00&0.00&0.00&0&0&0\\
18.2025&-0.2900&SDSSJ01128-0017&0.018&2&15.12&14.66&14.43&52963&15.16&14.74&14.35&0&2&3\\
18.7025&-0.4958&UGC793&0.034&1.5&16.83&16.48&16.28&53272&16.91&16.54&16.23&0&2&3\\
19.0300&33.0894&MARK1&0.016&2&15.92&15.95&15.78&0&0.00&0.00&0.00&2&2&41\\
20.8383&-1.9767&UM319&0.016&2&17.01&16.63&16.37&54770&17.08&16.74&16.35&0&2&2\\
21.1117&33.7994&NGC513&0.019&1&15.61&15.16&14.85&53263&15.71&15.26&14.85&1&2&5\\
21.3808&32.1364&MARK993&0.017&1.5&15.50&15.10&14.81&53263&15.63&15.21&14.82&0&0&0\\
21.8854&19.1789&MARK359&0.017&1&15.27&15.02&14.86&54848&15.37&15.17&14.85&0&2&5\\
22.2146&2.4464&UM105&0.030&2&16.44&16.09&15.72&54742&16.56&16.10&15.73&0&0&0\\
23.3800&35.6681&MARK1157&0.015&1&15.56&15.38&15.02&0&0.00&0.00&0.00&0&0&0\\
25.9908&2.3497&MARK573&0.017&1&15.05&14.90&14.60&54742&15.12&14.95&14.62&0&2&4\\
26.1267&17.1025&IIIZw35&0.027&2&16.25&15.83&15.51&54741&16.37&15.95&15.58&0&0&0\\
\hline
\end{tabular}
\begin{tablenotes}
\item[a]$^a$ Optical colour selection flag where 1 indicates possible narrowing spectra and 2 indicates possible broadening spectra.

\item[b]$^b$ CLAGN spectra flag where: 0 No archival spectra online, 1 Varying in archival and/or WiFeS spectra, 2 Only one archival spectrum found and no WiFeS, 3 Two or more archival spectra, not varying in WiFeS and archival spectra, 4 No archival spectra found from the last 10 years, but shows signs of varying before then and 5 No archival spectra found from the past 10 years, does not show signs of varying before then.

\item[c]$^c$ If Spectra Notes is 0, then this is also 0. 
1 WiFeS
2 6dFGS
3 SDSS
4 S7
5 BASS
6 2dFGRS
7 \citet{Ho1995} 
8 MaNGA
9 MUSE
10 \citet{Phi1983} 
11 \citet{Goo1983} 
12 \citet{Pen1984} 
13 \citet{Ost1985} 
14 \citet{Ber1986} 
15 \citet{Ver1986} 
16 \citet{Rud1988} 
17 \citet{Mor1988} 
18 \citet{Sab1989} 
19 \citet{Gre1991} 
20 \citet{Ken1992} 
21 \citet{Tra19922} 
22 \citet{Gri1992} 
23 \citet{Dur1994} 
24 \citet{Kim1995} 
25 \citet{Goo1995} 
26 \citet{Mor1996} 
27 \citet{Owe1996} 
28 \citet{Sca1996} 
29 \citet{Coz1997} 
30 \citet{Pie1998} 
31 \citet{Wei1999} 
32 \citet{Gon1999} 
33 \citet{Whi2000} 
34 2000UZC...C......0F
35 \citet{Rei2001} 
36 \citet{Ste2002} 
37 \citet{Ros2006} 
38 \citet{Mou2006} 
39 \citet{Lir2007} 
40 \citet{But2009} 
41 \citet{Sto2009} 
42 \citet{Tsa2009} 
43 \citet{Tri2010} 
44 \citet{Gav2013} 
45 \citet{Bar2015} 
46 \citet{Dop22015} 
47 \citet{Sch2016} 
48 \citet{Ram2016} 
49 \citet{Tho2017} 
50 \citet{Gre1997} 

\end{tablenotes}
\end{table}
\end{landscape}
}

\subsection{AGNs with 2 or more archival spectra}
To measure the completeness of our colour and flux selection criteria we inspected AGNs that have 2 or more archival spectra. Our main sources of spectra are WiFeS Siding Spring Southern Seyfert Spectroscopic Snapshot Survey \citep[S7;][]{Dop22015}, SDSS, 6dFGS and \citet{Ho1995}. The date ranges for these spectra sources are as follows: S7 spectra were observed from 2013-2016, SDSS spectra were observed between 2000-2019 (where DR16 spectra were observed through 2019), 6dFGS spectra were observed from 2001-2009 and \citet{Ho1995} spectra were observed between 1984-1990. Of all the $z$ $<$0.04 AGNs in our sample, 21$\%$ do not have readily available spectra online and 52$\%$ have only one available spectrum and 27$\%$ have multiple archival spectra. As we are refining our CLAGN candidates by selecting candidates where the archival spectra already show some signs of change, there is the potential to miss CLAGNs where the spectra have changed after the last archival spectra was taken and this will decrease our completeness. Of the AGNs that were not identified by our CLAGN candidate selection methods and have multiple archival spectra, we identified AGNs that appeared to have some small variations in the spectra, however these variations were not significant enough to require further investigation (i.e. changes in spectral class were 0.1 or less). Thus, while our three candidate selection criteria select many AGNs that are not CLAGNs, our selection criteria are not missing a large number of nearby CLAGNs.


\section{Spectroscopic follow-up}
\label{sec:observations}

\begin{table*} 
\caption{CLAGN candidates that were observed with WiFeS between 2018 July and 2019 March.}
\label{tbl:observed}
 \scriptsize
\begin{tabular}{cccccc}
\hline
 2MASX ID &Name& redshift & Initial type & WiFeS classification & MJD\\ [0.5ex]
    \hline
J03252538-0608380 & Mrk 609  & 0.0345& 1.9,2 & 1.9 & 58375\\
J03301327-0532363$^a$ & NGC 1346 & 0.0135 &1.8 & 2 & 58457\\
J05521140-0727222 & NGC 2110 & 
0.0078 & 1 & 1 &58491\\
J08044636+1046363 & UGC 04211 & 0.0344 & 1 &1 & 58548\\
J08353877-0405172$^b$ & NGC 2617 & 0.0142 & 1.8 & 1 & 58491\\
J10445172+0635488 & NGC 3362 & 
0.0277 & 2 &2&58491\\
J13254405-2950012 & NGC 5135 &  0.0137 & 2 &2 &58549\\
J13311382-2524096 & ESO 509- G 038 &  0.0260 & 1 &1 &58548\\
J13352457+0124376 & NGC 5227 & 0.0175 & 2 &2&58548\\
J15461637+0224558 & NGC 5990 &
0.0128 & 2 &2&58549\\
J20075129-1108346$^a$ & &  0.03 &2 & 1.8 & 58308\\
J21141259+0210406 & IC 1368 &  0.0130 & 2&2 &58310\\ 
J21522605-0810248 &  &  0.0348 & 2 &2&58375 \\
J22590139-2531423 & ESO 535- G 001 &  0.0303 & 2 &2& 58309\\
J22364648-1232426 & Mrk 915 & 0.0241 & 2& 1.9 & 58724\\

\hline
\end{tabular}
\begin{tablenotes}
\item[a]$^a$ New CLAGN that we have identified
\item[b] $^b$ Known CLAGN
\end{tablenotes}
\end{table*}

Once candidates were identified using the colour and flux criteria discussed in Sections~\ref{sec:opticalcolour},~\ref{sec:opticalflux} and \ref{sec:IRflux}, we inspected the archival spectra of these objects in order to identify potential CLAGNs without obtaining new spectra. It should be noted that we did not follow up candidates where archival spectra from the past two decades showed no variability (irrespective of the selection criteria). 

We used the Wide Field Spectrograph \citep[WiFeS,][]{Dop2010} Integral Field Unit (IFU) on the Australian National University's 2.3~m telescope at Siding Spring to obtain new spectra of our candidates to confirm that they are indeed CLAGN. WiFeS has a field of view of 25$^{\prime\prime}$ $\times$ 38$^{\prime\prime}$, divided into 950 spaxels. The advantage of using an IFU for follow-up observations of candidates is that extraction aperture size can be matched to previous observations \citep[e.g. the $7^{\prime\prime}$ fibre of 6dF,][]{Jon2009}. The wavelength coverage is  $3800 - 9200$~\AA, which spans the H$\alpha$, H$\beta$ and OIII lines at z $<$ 0.04, and we expect H$\alpha$ to show the most clear signs of change in CLAGNs.

We observed in nod-and-shuffle mode taking at least three frames with 60s on object, 60s on sky and 10 cycles per frame. This results in 40 min on object, 40 min on sky, and $\sim$ 15 min on overheads including telescope nod time, guide star re-acquisition and CCD readout. In total we allow $\sim$100 mins per galaxy. Our observations were taken between July 2018 and 2019 March. We reduced our data with PyWiFeS, Python-based pipeline \citep{Chi2014}. We observed 15 CLAGN candidates over multiple observing sessions, the AGNs are presented in Table~\ref{tbl:observed}.


\section{Spectral variability and new CLAGNs}
\label{sec:results}

We next present the spectra of the new CLAGNs that we have identified using the selection criteria mentioned in Sections~\ref{sec:opticalcolour},~\ref{sec:opticalflux} and \ref{sec:IRflux}, where we compare archival spectra of the galaxies with spectra taken using WiFeS. We plot archival spectra, where available, with the spectra taken using WiFeS to display the change. Multiplicative scaling has been applied so that the continuum spectra agree, highlighting changes in the emission lines. We match apertures of the multiple spectra of the new CLAGNs and AGNs with varying spectra below (where possible). It should be noted however, that the features in our observed WiFeS spectra displayed their respective broad line and narrow line features irrespective of the extraction aperture used.

\subsection{2MASX J20075129-1108346}
\begin{figure*} 
    \centering
    \includegraphics[width=1 \textwidth]{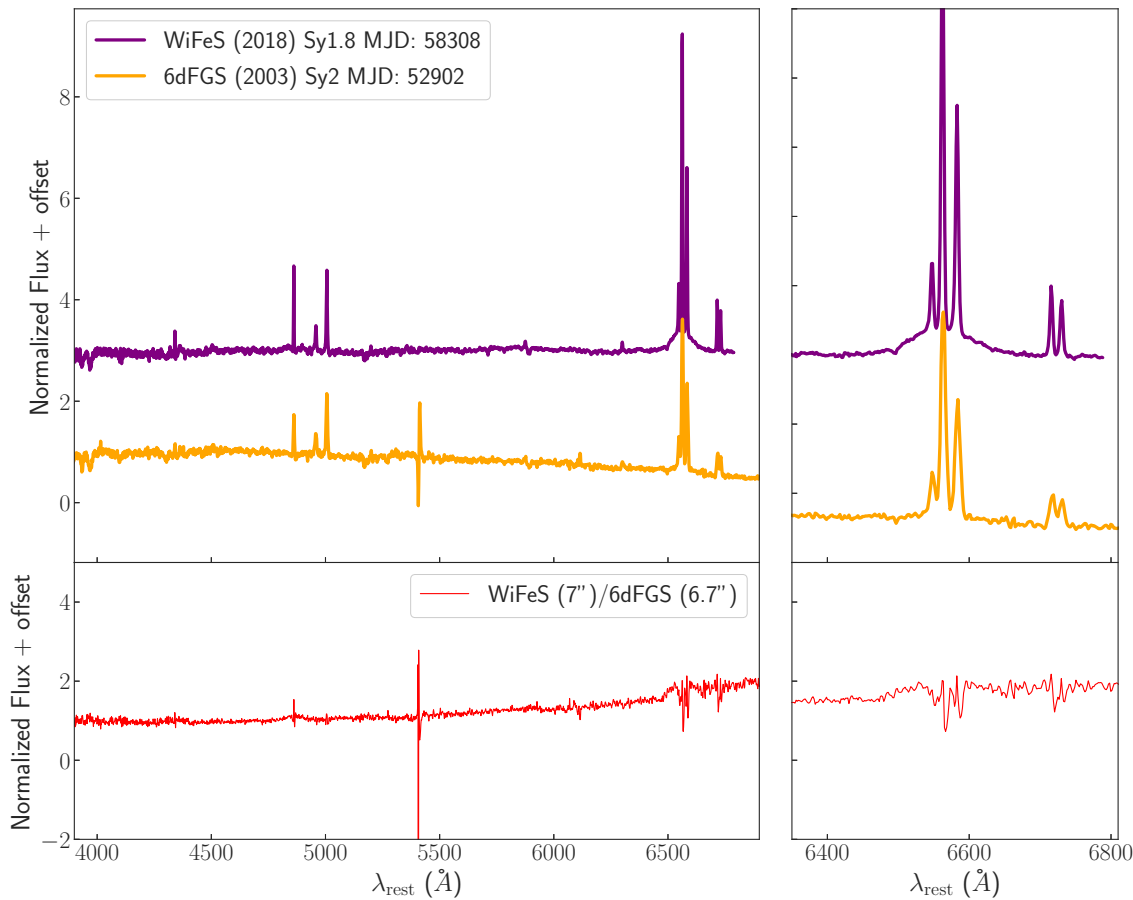}
    \caption{2MASX J20075129-1108346 is a new CLAGN which was a type 2 in the 6dFGS spectrum observed in 2003, and was determined to be a broad type 1.8 in the WiFeS spectrum taken in 2018. These classifications were made in accordance to the descriptions in \citet{Ost1981}.}
    \label{fig:J200}
\end{figure*}

2MASX J20075129-1108346 was classified as a type 1.9 by \citet{Ver2010}, however it has SkyMapper $r$ - $i$ $<$ 0.35 mag, which according to the optical colour selection criteria we adopt in Section~\ref{sec:opticalcolour}, suggests that the spectra of this AGN may have broadened. The 6dFGS spectrum in Figure~\ref{fig:J200} is consistent with a type 2 as the spectra contains only narrow line components. Our 2018 WiFeS spectrum shows broadline components (irrespective of the extraction aperture used) and we classify it as a type 1.8. As 2MASX J20075129-1108346 changes from type 2 to type 1.8. This change meets our criteria for a CLAGN.

\subsection{Mrk 609}

\begin{figure*} 
    \centering
    \includegraphics[width=1 \textwidth]{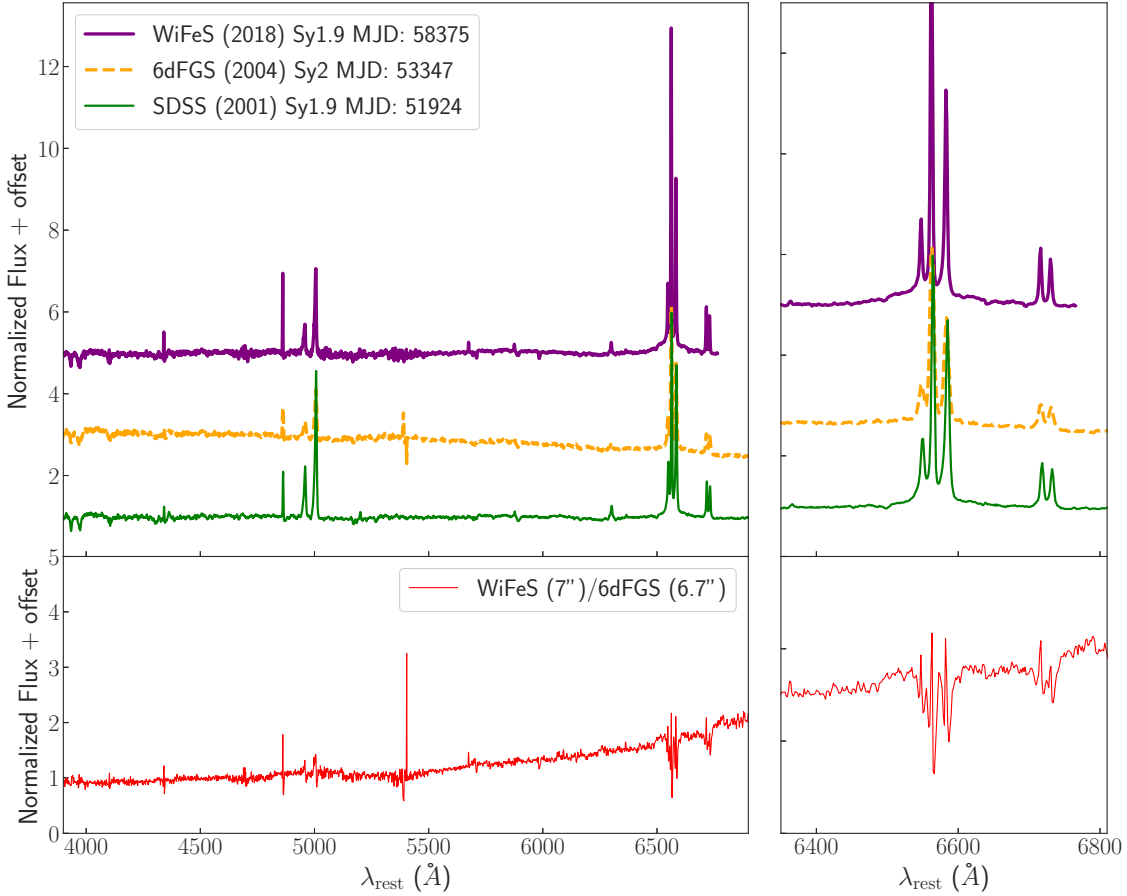}
    \caption{Mrk 609 has varying spectra and was identified using optical colour selection. The SDSS spectrum is consistent with a type 1.9 where the 6dFGS spectrum is completely narrow indicating it is a type 2. We classify Mrk 609 as type 1.9 using our WiFeS spectrum, and thus the changes in spectral class are insufficient to meet our CLAGN criterion.}
    \label{fig:mrk609}
\end{figure*}

Mrk 609 is classified by \citet{Ver2010} as a type 1.8 using the spectra from \citet{Rud1988}, but has $r$ - $i$ $<$ 0.35 mag in SkyMapper, indicating according to our colour selection criteria that it may have broadline components. Mrk 609 was one of the first Seyferts to be classified as an intermediate type. Mrk 609 was classified as a type 1.8 by \citet{Ost1981}. \citet{Rud1988} note that the spectral lines were inconsistent with a type 1 Seyfert, i.e. it lacks broad lines. \citet{Tri2010} report small variability in Mrk 609 spectra. They note the absence of broadline components in their observed spectra which they classified as type 2, although prior observations of Mrk 609, including \citet{Ost1981}, note broad H$\alpha$ components. As the \citet{Rud1988} spectral classification is inconclusive, we use the SDSS and 6dFGS spectra in Figure~\ref{fig:mrk609} as the baseline for determining whether Mrk 609 is changing spectral type. We classify the SDSS and 6dFGS spectra as type 1.9 and type 2, respectively, in accordance with the criteria outlined by \citet{Ost1981}. Our WiFeS spectra is consistent with that of a type 1.9 and is similar to that of SDSS indicating that Mrk 609 changed spectral type between 2001 and 2018 to a type 2, and has returned to being a type 1.9. This small variation in spectra is not consistent with our CLAGN criterion, but additional spectroscopy may reveal further changes in the spectral class of Mrk 609.

\subsection{Mrk 915}
 
\begin{figure*}
\begin{center}
\includegraphics[width=1\textwidth,angle=0]{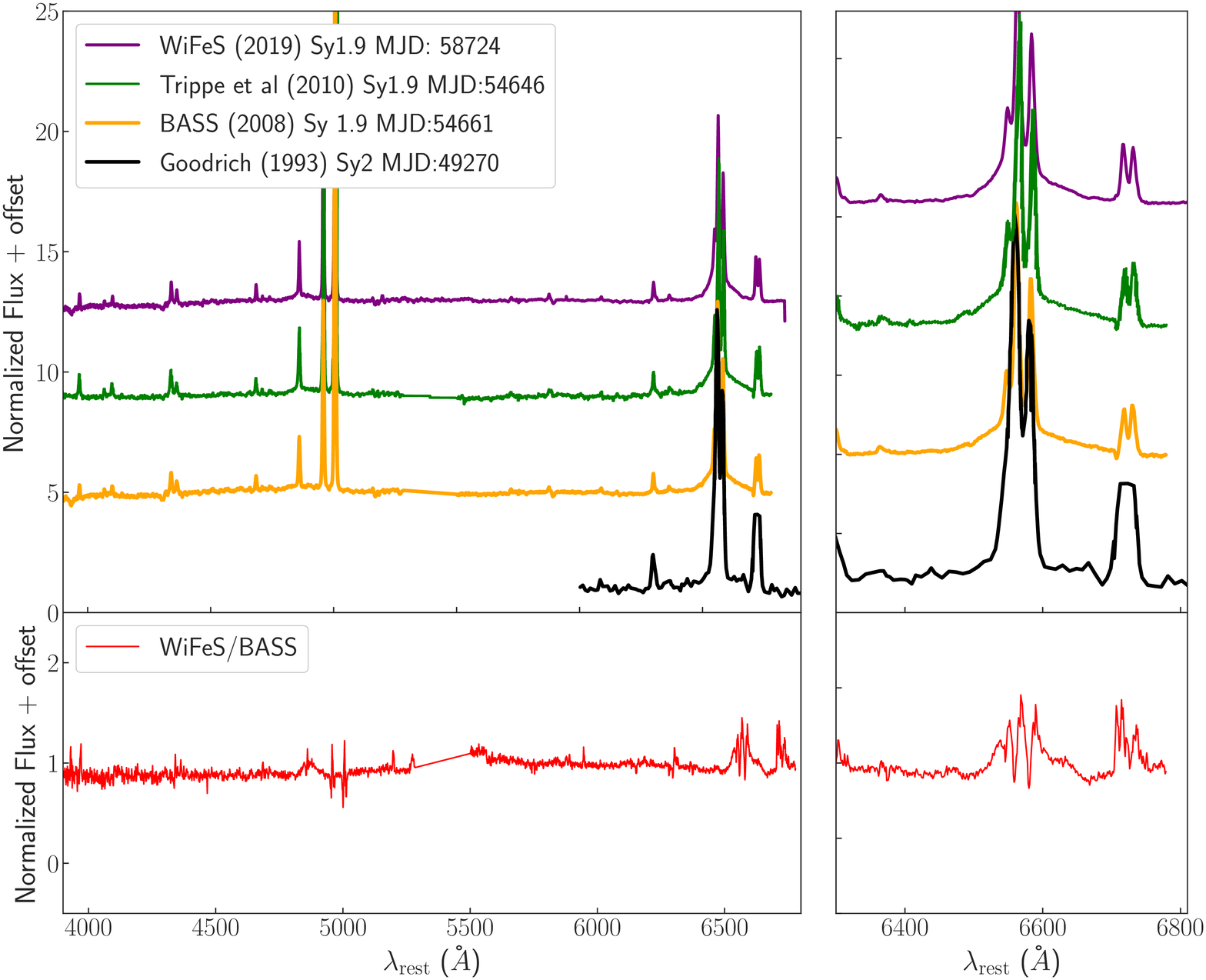}
\caption{Mrk 915 is a varying AGN that was identified using optical colour selection. We classify the 1993 \citep{Goo1995} spectrum as a type 2 as it contains only narrow lines and the BASS 2008 spectrum is consistent with a type 1.9. The WiFeS 2019 spectra is that of an type 1.9 \label{fig:Mrk915}. However, this change from type 2 to type 1.9 is not significant enough to meet out CLAGN criteria. Although this is the case, it is a good CLAGN candidate that will require further investigation.}
\end{center}
\end{figure*}

 Mrk 915 was classified as a type 1.8 by \citet{Ver2010} using the \citet{Dah1988} spectrum, but has Pan-STARRS $r$ - $i$ $<$ 0.25 mag. This colour, according to our Pan-STARRS colour selection criteria, suggests that the spectra of Mrk 915 has broadened. \citet{Goo1995} first reported the varying spectrum of Mrk 915, with a narrowing of emission lines between 1984 to 1993. \citet{Gia1996} also note a variation in the spectrum of Mrk 915, where they observed a broadening of the H$\alpha$ line between 1993 and 1994. While the 1993 spectrum in Figure~\ref{fig:Mrk915} is of relatively poor quality, it does not show the broadline component of subsequent spectra, and we conclude Mrk 915 as a type 2 Seyfert at the time. We classify the \textit{BAT AGN Spectroscopic Survey} \citep[BASS:][]{Kos2017} spectrum as type 1.9; the 2019 WiFeS spectrum in Figure~\ref{fig:Mrk915} is also consistent with a type 1.9. The H$\alpha$ line begins to broaden in the 2008 and 2010 spectrum (note: both spectra are from the same survey) and is broader still in the WiFeS 2019 spectrum. This variation in spectra from type 2 to type 1.9 does not meet our criteria for CLAGN. Although this may be the case, it is a good candidate CLAGN and further observations are needed.

\subsection{NGC 1346}

\begin{figure*} 
\begin{center}
\includegraphics[width=1\textwidth,angle=0]{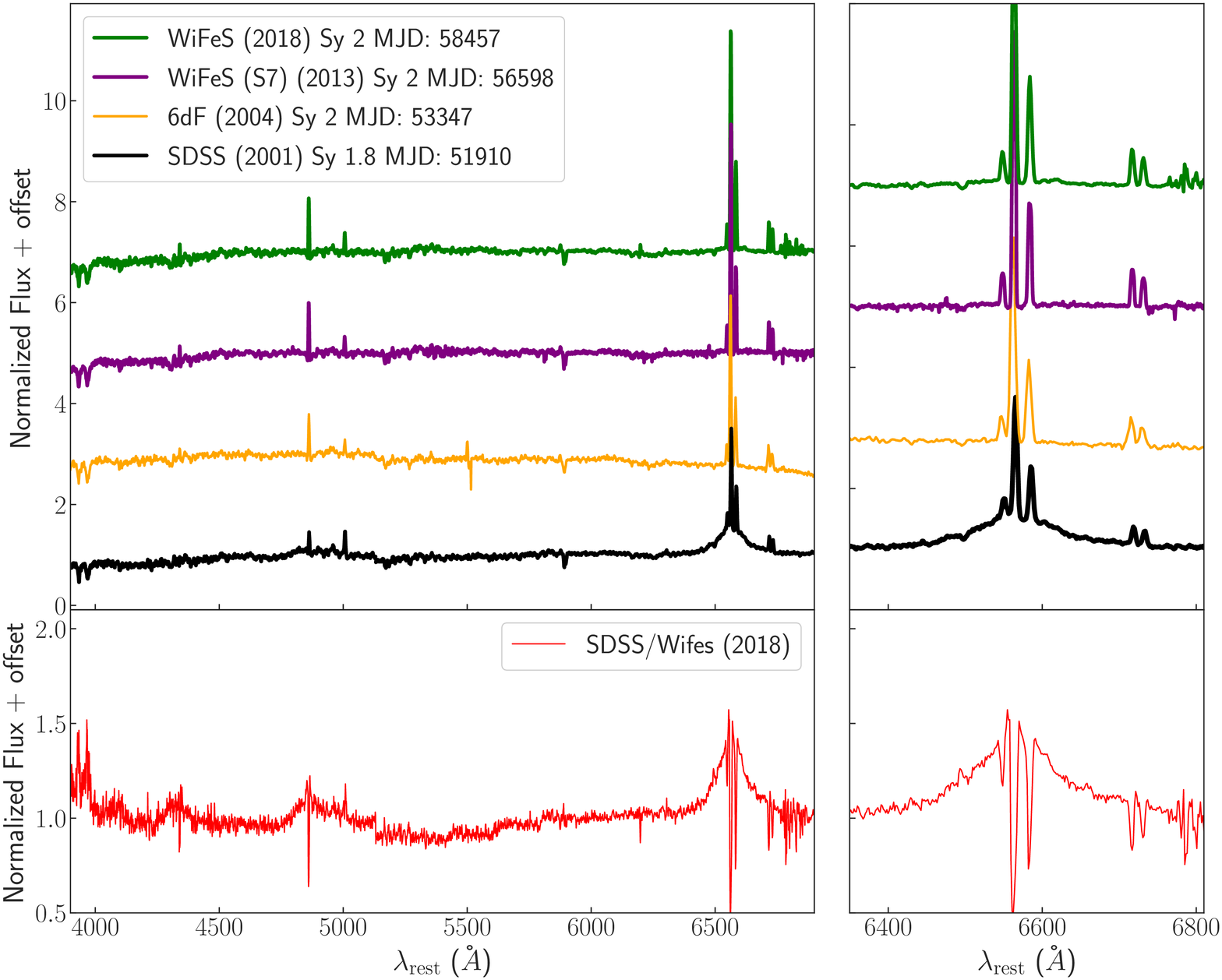}
\caption{Spectra of CLAGN NGC~1346, with the SDSS spectrum revealing broad ${\rm H\alpha}$ consistent with a type 1.8, while subsequent archival 6dFGS and WiFeS spectra, and our new WiFeS spectra, indicate a Seyfert 2 \citep{me}. \label{fig:NGC1346}}
\end{center}
\end{figure*}

NGC 1346 is a newly discovered CLAGN. We identified NGC 1346 as a broad-line AGN with unusually red colours with the SDSS and Pan-STARRS photometry and we designated it as a CLAGN via visual inspection of spectra from SDSS, 6dFGS and S7. NGC 1346 was classified as a Seyfert 1 galaxy by \citet{Ver2003} using the SDSS spectra in Figure~\ref{fig:NGC1346}. We classify this spectrum as a type 1.8 according to the definitions outlined by \citet{Ost1981}. The spectrum from SDSS (taken in 2001) showed a significant broad line component, however the 2004 December 6dFGS \citep{Jon2009} spectrum contains only narrow emission lines. The S7 spectrum of NGC 1346 and the WiFeS 2018 spectrum showed only narrow lines. Therefore, NGC 1346 was a type 2 prior to 2004 and it changed spectral type between 2001 and 2004. We use infrared photometry to investigate why this AGN is changing spectral type. 

To determine if a varying hot dust component of NGC 1346 could be responsible for the change in spectral type we use NEOWISE photometry and compared 2MASS photometry with recent targeted InfraRed Survey Facility \citep[IRSF;][]{Nag2012} photometry. The NEOWISE photometry was taken between 2014 and 2018 with a cadence of 6 months, and was thus taken after the change in spectral type. We measure a change in NEOWISE photometry of 0.11 mag; this is not a significant change and is not considered high enough to suggest a change in spectral type. This is because the NEOWISE survey has data from 2014 onward, and as suggested by the spectra, NGC 1346 had already changed spectral types by then. 

We find that NGC 1346 has faded by 0.82 mag in the $K_s$-band between 1998 and 2018, where we measure photometry from 2MASS and IRSF respectively. IR wavelength is sensitive to emission from warm dust attributed to the torus, as we measure a change in the IR photometry, this indicates that the change in spectral type we measure is a result of changes in the torus and not due to a simple obscuring cloud crossing the line of sight.

\section{Conclusions}
We have conducted a systematic survey for CLAGNs by identifying candidates using optical and infrared photometry from SkyMapper, Pan-STARRS, SDSS and NEOWISE. Using SkyMapper, we select type 1s with $r-i$ $>$ 0.53 mag and type 2s with $r$ - $i$ $<$ 0.35 mag and using Pan-STARRS we select type 1s with $r-i$ $>$ 0.43 mag and type 2s with $r$ - $i$ $<$ 0.25 mag. We also select candidates with optical $r$-band flux where type 1.8s, 1.9s and type 2s had $\Delta$m$_r$ $>$ 0.2, and search for AGNs with variability in \textit{WISE} W1(3.4 $\mu$m) (type 1.8s, 1.9s and type 2s where $\Delta$W1 $>$ 0.3 mag). Identifying candidates using optical colour selection provided the largest number of plausible candidates, with our new CLAGNs being selected in this manner. While this is the case, this selection criteria also produced the largest number of contaminants. The optical flux variability selection did not identify any new candidates, however it identified NGC 2617 as a candidate, showing it is a plausible method to identify CLAGNs.

Using NEOWISE W1 (3.4~$\mu$m) photometry we find majority of type 1s and type 2s have exhibited $>$ 0.3 mag and $<$ 0.3 mag, respectively, of variability during 2014 - 2018. While this allowed us to select type 2s displaying variability of $>$ 0.3 mag as CLAGN candidates, in practice all of the candidates selected appear to be (on the basis of archival spectra) misclassified broad line AGN. Thus \textit{WISE} W1 variability didn't prove useful for identifying changing-look Seyferts, but it could work with cleaner input catalogues and it has been used to identify changing-look quasars \citep[e.g.,][]{Guo22016,Ros2018,Ste2018}. 

Using our optical colour selection method we were able to identify four AGNs with varying spectra. 2MASX J20075129-110834 and NGC 1346 are new CLAGNs that were identified in this work using optical colour selection. Mrk 915 and Mrk 609 have varying spectra which do not meet our criteria for CLAGN and only have a small change from type 2 to type 1.9 and type 1.9 to type 2, respectively. These AGNs remain CLAGN candidates and additional followup spectroscopy may reveal further changes in their spectral types. 46$\%$ of candidates selected using this method either didn't have archival spectra at all, didn't have archival spectra from the last 10 years or only had one archival spectrum. Extrapolating this, we can estimate that we have only identified 54$\%$ of possible CLAGNs in this sample due to lack of spectra. The optical colour selection method also only identifies $\approx$ 50 $\%$ of known CLAGN. 

We note that as we refined our CLAGN candidates by selecting candidates where the archival spectra already showed signs of change, we may have missed CLAGNs that may have changed after the last archival spectra was taken and/or changed type relatively briefly (i.e. $<$ 2 years). To estimate the number of CLAGN candidates that could have changed spectral type rapidly or briefly, we need an estimate of the numbers of CLAGNs as a function of the timescale of variability. 1ES 1927+654 (z $=$ 0.017) is a CLAGN that changed from type 2 to type 1, where the change lasted for 11 months \citep{Tra2019}. For our candidates selected via optical and infrared flux changes the variability is on timescales of a decade to 6 months respectively. Optical flux variability will miss CLAGNs that vary only briefly from their usual state, and while NEOWISE has the cadence to detect such candidates the NEOWISE variability of 1ES 1927+654 remained < 0.2 mag (although the TDE produced $\sim$ 2 mags brightening in the optical) and thus it wasn’t flagged as a WISE selected CLAGN candidate.  We also note that clouds of dust moving across the line of sight may not significantly impact the NEOWISE photometry as this can occur on relatively short time scales \citep{Guo22016}, thus this subclass of CLAGN could be underrepresented in our sample. As such we have a lower limit of $\approx$ 18 CLAGN as z $<$ 0.04, which includes 2 new CLAGNS discovered in this work, 10 previously known CLAGNs that have varied between 1998 and 2015, and as many as 6 CLAGNs that missed our candidate selection methods or didn't have archival spectra.


\section*{Acknowledgements}
The authors would like to thank the anonymous referee for their helpful comments which improved the manuscript overall.

The authors wish to thank the staff at ANU 2.3m telescope and the WiFeS instrument for their technical support. We would also like to thank the ANU telescope time allocation committee for supporting this project and the observations in this paper. 

Michelle E. Cluver is a recipient of an Australian Research Council Future Fellowship (project number FT170100273) funded by the Australian Government.

John R. Lucey was supported by the Science and Technology Facilities Council through the Durham Astronomy Consolidated Grants ST/P000541/1 and ST/T000244/1.

The national facility capability for SkyMapper has been funded through ARC LIEF grant LE130100104 from the Australian Research Council, awarded to the University of Sydney, the Australian National University, Swinburne University of Technology, the University of Queensland, the University of Western Australia, the University of Melbourne, Curtin University of Technology, Monash University and the Australian Astronomical Observatory. SkyMapper is owned and operated by The Australian National University's Research School of Astronomy and Astrophysics. The survey data were processed and provided by the SkyMapper Team at ANU. The SkyMapper node of the All-Sky Virtual Observatory (ASVO) is hosted at the National Computational Infrastructure (NCI). Development and support the SkyMapper node of the ASVO has been funded in part by Astronomy Australia Limited (AAL) and the Australian Government through the Commonwealth's Education Investment Fund (EIF) and National Collaborative Research Infrastructure Strategy (NCRIS), particularly the National eResearch Collaboration Tools and Resources (NeCTAR) and the Australian National Data Service Projects (ANDS).

The Pan-STARRS1 Surveys (PS1) and the PS1 public science archive have been made possible through contributions by the Institute for Astronomy, the University of Hawaii, the Pan-STARRS Project Office, the Max-Planck Society and its participating institutes, the Max Planck Institute for Astronomy, Heidelberg and the Max Planck Institute for Extraterrestrial Physics, Garching, The Johns Hopkins University, Durham University, the University of Edinburgh, the Queen's University Belfast, the Harvard-Smithsonian Center for Astrophysics, the Las Cumbres Observatory Global Telescope Network Incorporated, the National Central University of Taiwan, the Space Telescope Science Institute, the National Aeronautics and Space Administration under Grant No. NNX08AR22G issued through the Planetary Science Division of the NASA Science Mission Directorate, the National Science Foundation Grant No. AST-1238877, the University of Maryland, Eotvos Lorand University (ELTE), the Los Alamos National Laboratory, and the Gordon and Betty Moore Foundation.

SDSS-IV is managed by the Astrophysical Research Consortium for the Participating Institutions of the SDSS Collaboration including the Brazilian Participation Group, the Carnegie Institution for Science, Carnegie Mellon University, the Chilean Participation Group, the French Participation Group, Harvard-Smithsonian Center for Astrophysics, Instituto de Astrof\'isica de Canarias, The Johns Hopkins University, Kavli Institute for the Physics and Mathematics of the Universe (IPMU) / University of Tokyo, the Korean Participation Group, Lawrence Berkeley National Laboratory, Leibniz Institut f\"ur Astrophysik Potsdam (AIP), Max-Planck-Institut f\"ur Astronomie (MPIA Heidelberg), Max-Planck-Institut f\"ur Astrophysik (MPA Garching), Max-Planck-Institut f\"ur Extraterrestrische Physik (MPE), National Astronomical Observatories of China, New Mexico State University, New York University, University of Notre Dame, Observat\'ario Nacional / MCTI, The Ohio State University, Pennsylvania State University, Shanghai Astronomical Observatory, United Kingdom Participation Group, Universidad Nacional Aut\'onoma de M\'exico, University of Arizona, University of Colorado Boulder, University of Oxford, University of Portsmouth, University of Utah, University of Virginia, University of Washington, University of Wisconsin, Vanderbilt University, and Yale University.

This publication makes use of data products from the Wide-field Infrared Survey Explorer, which is a joint project of the University of California, Los Angeles, and the Jet Propulsion Laboratory/California Institute of Technology, funded by the National Aeronautics and Space Administration. Funding for the Sloan Digital Sky Survey IV has been provided by the Alfred P. Sloan Foundation, the U.S. Department of Energy Office of Science, and the Participating Institutions. SDSS-IV acknowledges support and resources from the Center for High-Performance Computing at the University of Utah. The SDSS web site is www.sdss.org.

This research has made use of the NASA/IPAC Extragalactic Database (NED) which is operated by the Jet Propulsion Laboratory, California Institute of Technology, under contract with the National Aeronautics and Space Administration and the HyperLeda database (http://leda.univlyon1.fr). This research made use of Astropy, a community developed core Python package for Astronomy (Astropy Collaboration, 2018).

\section*{Data Availability}
The data underlying this article are available in the article and in its online supplementary material.









\bibliographystyle{mnras}
\bibliography{References}

\begin{thebibliography}{}
\makeatletter
\relax
\def\mn@urlcharsother{\let\do\@makeother \do\$\do\&\do\#\do\^\do\_\do\%\do\~}
\def\mn@doi{\begingroup\mn@urlcharsother \@ifnextchar [ {\mn@doi@}
  {\mn@doi@[]}}
\def\mn@doi@[#1]#2{\def\@tempa{#1}\ifx\@tempa\@empty \href
  {http://dx.doi.org/#2} {doi:#2}\else \href {http://dx.doi.org/#2} {#1}\fi
  \endgroup}
\def\mn@eprint#1#2{\mn@eprint@#1:#2::\@nil}
\def\mn@eprint@arXiv#1{\href {http://arxiv.org/abs/#1} {{\tt arXiv:#1}}}
\def\mn@eprint@dblp#1{\href {http://dblp.uni-trier.de/rec/bibtex/#1.xml}
  {dblp:#1}}
\def\mn@eprint@#1:#2:#3:#4\@nil{\def\@tempa {#1}\def\@tempb {#2}\def\@tempc
  {#3}\ifx \@tempc \@empty \let \@tempc \@tempb \let \@tempb \@tempa \fi \ifx
  \@tempb \@empty \def\@tempb {arXiv}\fi \@ifundefined
  {mn@eprint@\@tempb}{\@tempb:\@tempc}{\expandafter \expandafter \csname
  mn@eprint@\@tempb\endcsname \expandafter{\@tempc}}}

\bibitem[\protect\citeauthoryear{{Abazajian}, {Adelman-McCarthy},
  {Ag{\"u}eros}, {Allam}  \& et al.}{{Abazajian} et~al.}{2004}]{Aba2004}
{Abazajian} K.,  {Adelman-McCarthy} J.~K.,  {Ag{\"u}eros} M.~A.,  {Allam}
  S.~S.,   et al. 2004, \mn@doi [Astronomical Journal] {10.1086/421365}, \href
  {http://adsabs.harvard.edu/abs/2004AJ....128..502A} {128, 502}

\bibitem[\protect\citeauthoryear{{Afanasiev}, {Popovi{\'c}}, {Shapovalova},
  {Borisov}  \& {Ili{\'c}}}{{Afanasiev} et~al.}{2014}]{Afa2014}
{Afanasiev} V.~L.,  {Popovi{\'c}} L.~{\v{C}}.,  {Shapovalova} A.~I.,  {Borisov}
  N.~V.,   {Ili{\'c}} D.,  2014, \mn@doi [\mnras] {10.1093/mnras/stu231}, \href
  {https://ui-adsabs-harvard-edu.ezproxy.lib.monash.edu.au/abs/2014MNRAS.440..519A}
  {440, 519}

\bibitem[\protect\citeauthoryear{{Ag{\'{\i}}s-Gonz{\'a}lez}, {Bagnulo},
  {Hutsem{\'e}kers}, {Montesinos}, {Miniutti}  \&
  {Sanfrutos}}{{Ag{\'{\i}}s-Gonz{\'a}lez} et~al.}{2017}]{Agi2017}
{Ag{\'{\i}}s-Gonz{\'a}lez} B.,  {Bagnulo} S.,  {Hutsem{\'e}kers} D.,
  {Montesinos} B.,  {Miniutti} G.,   {Sanfrutos} M.,  2017, in {Arribas} S.,
  {Alonso-Herrero} A.,  {Figueras} F.,  {Hern{\'a}ndez-Monteagudo} C.,
  {S{\'a}nchez-Lavega} A.,   {P{\'e}rez-Hoyos} S.,  eds, Highlights on Spanish
  Astrophysics IX. pp 275--275

\bibitem[\protect\citeauthoryear{{Antonucci}}{{Antonucci}}{1993}]{Ant1993}
{Antonucci} R.,  1993, \mn@doi [\araa] {10.1146/annurev.aa.31.090193.002353},
  \href {http://adsabs.harvard.edu/abs/1993ARA%26A..31..473A} {31, 473}

\bibitem[\protect\citeauthoryear{{Barth} et~al.,}{{Barth}
  et~al.}{2015}]{Bar2015}
{Barth} A.~J.,  et~al., 2015, \mn@doi [\apjs] {10.1088/0067-0049/217/2/26},
  \href {https://ui.adsabs.harvard.edu/abs/2015ApJS..217...26B} {217, 26}

\bibitem[\protect\citeauthoryear{{Bergvall}, {Johansson}  \&
  {Olofsson}}{{Bergvall} et~al.}{1986}]{Ber1986}
{Bergvall} N.,  {Johansson} L.,   {Olofsson} K.,  1986, \aap, \href
  {https://ui.adsabs.harvard.edu/abs/1986A&A...166...92B} {166, 92}

\bibitem[\protect\citeauthoryear{{Bianchi}, {La Franca}, {Matt}, {Guainazzi},
  {Jimenez Bail{\'o}n}, {Longinotti}, {Nicastro}  \& {Pentericci}}{{Bianchi}
  et~al.}{2008}]{Bia2008}
{Bianchi} S.,  {La Franca} F.,  {Matt} G.,  {Guainazzi} M.,  {Jimenez
  Bail{\'o}n} E.,  {Longinotti} A.~L.,  {Nicastro} F.,   {Pentericci} L.,
  2008, \mn@doi [Monthly Notices of the Royal Astronomical Society]
  {10.1111/j.1745-3933.2008.00521.x}, \href
  {http://adsabs.harvard.edu/abs/2008MNRAS.389L..52B} {389, L52}

\bibitem[\protect\citeauthoryear{{Braito}, {Reeves}, {Bianchi}, {Nardini}  \&
  {Piconcelli}}{{Braito} et~al.}{2017}]{Bra2017}
{Braito} V.,  {Reeves} J.~N.,  {Bianchi} S.,  {Nardini} E.,   {Piconcelli} E.,
  2017, \mn@doi [\aap] {10.1051/0004-6361/201630322}, \href
  {https://ui.adsabs.harvard.edu/abs/2017A&A...600A.135B} {600, A135}

\bibitem[\protect\citeauthoryear{{Burtscher} et~al.,}{{Burtscher}
  et~al.}{2015}]{Bur2015}
{Burtscher} L.,  et~al., 2015, \mn@doi [\aap] {10.1051/0004-6361/201525817},
  \href {https://ui.adsabs.harvard.edu/abs/2015A&A...578A..47B} {578, A47}

\bibitem[\protect\citeauthoryear{{Buttiglione}, {Capetti}, {Celotti}, {Axon},
  {Chiaberge}, {Macchetto}  \& {Sparks}}{{Buttiglione} et~al.}{2009}]{But2009}
{Buttiglione} S.,  {Capetti} A.,  {Celotti} A.,  {Axon} D.~J.,  {Chiaberge} M.,
   {Macchetto} F.~D.,   {Sparks} W.~B.,  2009, \mn@doi [\aap]
  {10.1051/0004-6361:200811102}, \href
  {https://ui.adsabs.harvard.edu/abs/2009A&A...495.1033B} {495, 1033}

\bibitem[\protect\citeauthoryear{{Canelo}, {Fria{\c c}a}, {Sales}, {Pastoriza}
  \& {Ruschel-Dutra}}{{Canelo} et~al.}{2018}]{Can2018}
{Canelo} C.~M.,  {Fria{\c c}a} A.~C.~S.,  {Sales} D.~A.,  {Pastoriza} M.~G.,
  {Ruschel-Dutra} D.,  2018, \mn@doi [Monthly Notices of the Royal Astronomical
  Society] {10.1093/mnras/stx3351}, \href
  {http://adsabs.harvard.edu/abs/2018MNRAS.475.3746C} {475, 3746}

\bibitem[\protect\citeauthoryear{Chambers et~al.,}{Chambers
  et~al.}{2019b}]{Cha2019}
Chambers K.~C.,  et~al., 2019b, The Pan-STARRS1 Surveys (\mn@eprint {arXiv}
  {1612.05560})

\bibitem[\protect\citeauthoryear{{Chambers} et~al.,}{{Chambers}
  et~al.}{2019a}]{Cha2016}
{Chambers} K.~C.,  et~al., 2019a, Transient Name Server Discovery Report, \href
  {https://arxiv.org/abs/1612.05560} {}

\bibitem[\protect\citeauthoryear{{Childress}, {Vogt}, {Nielsen}  \&
  {Sharp}}{{Childress} et~al.}{2014}]{Chi2014}
{Childress} M.~J.,  {Vogt} F. P.~A.,  {Nielsen} J.,   {Sharp} R.~G.,  2014,
  \mn@doi [\apss] {10.1007/s10509-013-1682-0}, \href
  {https://ui.adsabs.harvard.edu/abs/2014Ap&SS.349..617C} {349, 617}

\bibitem[\protect\citeauthoryear{{Coziol}, {Demers}, {Barneoud}  \&
  {Pena}}{{Coziol} et~al.}{1997}]{Coz1997}
{Coziol} R.,  {Demers} S.,  {Barneoud} R.,   {Pena} M.,  1997, \mn@doi [\aj]
  {10.1086/118372}, \href
  {https://ui.adsabs.harvard.edu/abs/1997AJ....113.1548C} {113, 1548}

\bibitem[\protect\citeauthoryear{{Cruz-Gonzalez}, {Carrasco}, {Serrano},
  {Guichard}, {Dultzin-Hacyan}  \& {Bisiacchi}}{{Cruz-Gonzalez}
  et~al.}{1994}]{Cru1994}
{Cruz-Gonzalez} I.,  {Carrasco} L.,  {Serrano} A.,  {Guichard} J.,
  {Dultzin-Hacyan} D.,   {Bisiacchi} G.~F.,  1994, \mn@doi [\apjs]
  {10.1086/192072}, \href
  {https://ui.adsabs.harvard.edu/abs/1994ApJS...94...47C} {94, 47}

\bibitem[\protect\citeauthoryear{{Dahari} \& {De Robertis}}{{Dahari} \& {De
  Robertis}}{1988}]{Dah1988}
{Dahari} O.,  {De Robertis} M.~M.,  1988, \mn@doi [\apjs] {10.1086/191273},
  \href {https://ui.adsabs.harvard.edu/abs/1988ApJS...67..249D} {67, 249}

\bibitem[\protect\citeauthoryear{Davis \& El-Abd}{Davis \&
  El-Abd}{2019}]{Dav2019}
Davis S.~W.,  El-Abd S.,  2019, \mn@doi [The Astrophysical Journal]
  {10.3847/1538-4357/ab05c5}, 874, 23

\bibitem[\protect\citeauthoryear{{Denney} et~al.,}{{Denney}
  et~al.}{2014}]{Den2014}
{Denney} K.~D.,  et~al., 2014, \mn@doi [\apj] {10.1088/0004-637X/796/2/134},
  \href {http://adsabs.harvard.edu/abs/2014ApJ...796..134D} {796, 134}

\bibitem[\protect\citeauthoryear{{Dexter} \& {Agol}}{{Dexter} \&
  {Agol}}{2011}]{Dex2011}
{Dexter} J.,  {Agol} E.,  2011, \mn@doi [The Astrophysical Journal]
  {10.1088/2041-8205/727/1/L24}, \href
  {http://adsabs.harvard.edu/abs/2011ApJ...727L..24D} {727, L24}

\bibitem[\protect\citeauthoryear{{Dopita} et~al.,}{{Dopita}
  et~al.}{2010}]{Dop2010}
{Dopita} M.,  et~al., 2010, \mn@doi [\apss] {10.1007/s10509-010-0335-9}, \href
  {https://ui.adsabs.harvard.edu/abs/2010Ap%26SS.327..245D} {327, 245}

\bibitem[\protect\citeauthoryear{{Dopita} et~al.,}{{Dopita}
  et~al.}{2014}]{Dop2014}
{Dopita} M.~A.,  et~al., 2014, \mn@doi [\aap] {10.1051/0004-6361/201423467},
  \href {https://ui.adsabs.harvard.edu/abs/2014A&A...566A..41D} {566, A41}

\bibitem[\protect\citeauthoryear{{Dopita} et~al.,}{{Dopita}
  et~al.}{2015}]{Dop22015}
{Dopita} M.~A.,  et~al., 2015, \mn@doi [\apjs] {10.1088/0067-0049/217/1/12},
  \href
  {https://ui-adsabs-harvard-edu.ezproxy.lib.monash.edu.au/abs/2015ApJS..217...12D}
  {217, 12}

\bibitem[\protect\citeauthoryear{{Durret}}{{Durret}}{1994}]{Dur1994}
{Durret} F.,  1994, \aaps, \href
  {https://ui.adsabs.harvard.edu/abs/1994A&AS..105...57D} {105, 57}

\bibitem[\protect\citeauthoryear{{Eisenstein} et~al.,}{{Eisenstein}
  et~al.}{2011}]{Eis2011}
{Eisenstein} D.~J.,  et~al., 2011, \mn@doi [\aj] {10.1088/0004-6256/142/3/72},
  \href {http://adsabs.harvard.edu/abs/2011AJ....142...72E} {142, 72}

\bibitem[\protect\citeauthoryear{{Eracleous} \& {Halpern}}{{Eracleous} \&
  {Halpern}}{2001}]{Era2001}
{Eracleous} M.,  {Halpern} J.~P.,  2001, \mn@doi [Astrophysical Journal]
  {10.1086/321331}, \href {http://adsabs.harvard.edu/abs/2001ApJ...554..240E}
  {554, 240}

\bibitem[\protect\citeauthoryear{{Flewelling} et~al.,}{{Flewelling}
  et~al.}{2016}]{Fle2016}
{Flewelling} H.~A.,  et~al., 2016, arXiv e-prints, \href
  {https://ui.adsabs.harvard.edu/abs/2016arXiv161205243F} {}

\bibitem[\protect\citeauthoryear{{Fosbury} et~al.,}{{Fosbury}
  et~al.}{1982}]{Fos1982}
{Fosbury} R.~A.~E.,  et~al., 1982, \mn@doi [\mnras] {10.1093/mnras/201.4.991},
  \href {https://ui.adsabs.harvard.edu/abs/1982MNRAS.201..991F} {201, 991}

\bibitem[\protect\citeauthoryear{{Fraquelli}, {Storchi-Bergmann}  \&
  {Binette}}{{Fraquelli} et~al.}{2000}]{Fra2000}
{Fraquelli} H.~A.,  {Storchi-Bergmann} T.,   {Binette} L.,  2000, \mn@doi
  [\apj] {10.1086/308621}, \href
  {https://ui.adsabs.harvard.edu/abs/2000ApJ...532..867F} {532, 867}

\bibitem[\protect\citeauthoryear{{Gaia Collaboration} et~al.,}{{Gaia
  Collaboration} et~al.}{2016}]{Pru2016}
{Gaia Collaboration} et~al., 2016, \mn@doi [\aap]
  {10.1051/0004-6361/201629272}, \href
  {http://adsabs.harvard.edu/abs/2016A%26A...595A...1G} {595, A1}

\bibitem[\protect\citeauthoryear{{Gavazzi}, {Consolandi}, {Dotti}, {Fossati},
  {Savorgnan}, {Gualandi}  \& {Bruni}}{{Gavazzi} et~al.}{2013}]{Gav2013}
{Gavazzi} G.,  {Consolandi} G.,  {Dotti} M.,  {Fossati} M.,  {Savorgnan} G.,
  {Gualandi} R.,   {Bruni} I.,  2013, \mn@doi [Astronomy and Astrophysics]
  {10.1051/0004-6361/201322016}, \href
  {http://adsabs.harvard.edu/abs/2013A$\%$26A...558A..68G} {558, A68}

\bibitem[\protect\citeauthoryear{{Georgantopoulos}, {Papadakis}, {Zezas}  \&
  {Ward}}{{Georgantopoulos} et~al.}{2004}]{Geo2004}
{Georgantopoulos} I.,  {Papadakis} I.,  {Zezas} A.,   {Ward} M.~J.,  2004,
  \mn@doi [\apj] {10.1086/423366}, \href
  {https://ui.adsabs.harvard.edu/abs/2004ApJ...614..634G} {614, 634}

\bibitem[\protect\citeauthoryear{{George} \& {Fabian}}{{George} \&
  {Fabian}}{1991}]{Geo1991}
{George} I.~M.,  {Fabian} A.~C.,  1991, \mn@doi [Monthly Notices of the Royal
  Astronomical Society] {10.1093/mnras/249.2.352}, \href
  {http://adsabs.harvard.edu/abs/1991MNRAS.249..352G} {249, 352}

\bibitem[\protect\citeauthoryear{{Gezari} et~al.,}{{Gezari}
  et~al.}{2017}]{Gez2017}
{Gezari} S.,  et~al., 2017, \mn@doi [\apj] {10.3847/1538-4357/835/2/144}, \href
  {https://ui-adsabs-harvard-edu.ezproxy.lib.monash.edu.au/abs/2017ApJ...835..144G}
  {835, 144}

\bibitem[\protect\citeauthoryear{{Giannuzzo} \& {Stirpe}}{{Giannuzzo} \&
  {Stirpe}}{1996}]{Gia1996}
{Giannuzzo} E.~M.,  {Stirpe} G.~M.,  1996, \aap, \href
  {https://ui.adsabs.harvard.edu/abs/1996A&A...314..419G} {314, 419}

\bibitem[\protect\citeauthoryear{{Gilli}, {Maiolino}, {Marconi}, {Risaliti},
  {Dadina}, {Weaver}  \& {Colbert}}{{Gilli} et~al.}{2000}]{Gil2000}
{Gilli} R.,  {Maiolino} R.,  {Marconi} A.,  {Risaliti} G.,  {Dadina} M.,
  {Weaver} K.~A.,   {Colbert} E.~J.~M.,  2000, \aap, \href
  {https://ui.adsabs.harvard.edu/abs/2000A&A...355..485G} {355, 485}

\bibitem[\protect\citeauthoryear{{Gon{\c{c}}alves}, {V{\'e}ron-Cetty}  \&
  {V{\'e}ron}}{{Gon{\c{c}}alves} et~al.}{1999}]{Gon1999}
{Gon{\c{c}}alves} A.~C.,  {V{\'e}ron-Cetty} M.~P.,   {V{\'e}ron} P.,  1999,
  \mn@doi [\aaps] {10.1051/aas:1999183}, \href
  {https://ui.adsabs.harvard.edu/abs/1999A&AS..135..437G} {135, 437}

\bibitem[\protect\citeauthoryear{{Goodrich}}{{Goodrich}}{1989}]{Goo1989}
{Goodrich} R.~W.,  1989, \mn@doi [The Astrophysical Journal] {10.1086/167384},
  \href {http://adsabs.harvard.edu/abs/1989ApJ...340..190G} {340, 190}

\bibitem[\protect\citeauthoryear{{Goodrich}}{{Goodrich}}{1995}]{Goo1995}
{Goodrich} R.~W.,  1995, \mn@doi [\apj] {10.1086/175256}, \href
  {https://ui-adsabs-harvard-edu.ezproxy.lib.monash.edu.au/abs/1995ApJ...440..141G}
  {440, 141}

\bibitem[\protect\citeauthoryear{{Goodrich} \& {Osterbrock}}{{Goodrich} \&
  {Osterbrock}}{1983}]{Goo1983}
{Goodrich} R.~W.,  {Osterbrock} D.~E.,  1983, \mn@doi [\apj] {10.1086/161052},
  \href {https://ui.adsabs.harvard.edu/abs/1983ApJ...269..416G} {269, 416}

\bibitem[\protect\citeauthoryear{{Greenawalt}, {Walterbos}  \&
  {Braun}}{{Greenawalt} et~al.}{1997}]{Gre1997}
{Greenawalt} B.,  {Walterbos} R.~A.~M.,   {Braun} R.,  1997, \mn@doi [\apj]
  {10.1086/304285}, \href
  {https://ui.adsabs.harvard.edu/abs/1997ApJ...483..666G} {483, 666}

\bibitem[\protect\citeauthoryear{{Gregory}, {Tifft}  \& {Cocke}}{{Gregory}
  et~al.}{1991}]{Gre1991}
{Gregory} S.~A.,  {Tifft} W.~G.,   {Cocke} W.~J.,  1991, \mn@doi [\aj]
  {10.1086/116019}, \href
  {https://ui.adsabs.harvard.edu/abs/1991AJ....102.1977G} {102, 1977}

\bibitem[\protect\citeauthoryear{{Guo} et~al.,}{{Guo} et~al.}{2016}]{Guo22016}
{Guo} H.,  et~al., 2016, \mn@doi [Astrophysical Journal]
  {10.3847/0004-637X/826/2/186}, \href
  {http://adsabs.harvard.edu/abs/2016ApJ...826..186G} {826, 186}

\bibitem[\protect\citeauthoryear{{Ho} \& {Kim}}{{Ho} \& {Kim}}{2009}]{Ho2009}
{Ho} L.~C.,  {Kim} M.,  2009, \mn@doi [\apjs] {10.1088/0067-0049/184/2/398},
  \href {https://ui.adsabs.harvard.edu/abs/2009ApJS..184..398H} {184, 398}

\bibitem[\protect\citeauthoryear{{Ho}, {Filippenko}  \& {Sargent}}{{Ho}
  et~al.}{1995}]{Ho1995}
{Ho} L.~C.,  {Filippenko} A.~V.,   {Sargent} W.~L.,  1995, \mn@doi [\apjs]
  {10.1086/192170}, \href
  {https://ui.adsabs.harvard.edu/abs/1995ApJS...98..477H} {98, 477}

\bibitem[\protect\citeauthoryear{Hon, Webster  \& Wolf}{Hon
  et~al.}{2020}]{Hon2020}
Hon W.~J.,  Webster R.,   Wolf C.,  2020, \mn@doi [Monthly Notices of the Royal
  Astronomical Society] {10.1093/mnras/staa1939}, 497, 192

\bibitem[\protect\citeauthoryear{H{\"o}nig et~al.,}{H{\"o}nig
  et~al.}{2013}]{Hon2013}
H{\"o}nig S.~F.,  et~al., 2013, \mn@doi [The Astrophysical Journal]
  {10.1088/0004-637x/771/2/87}, 771, 87

\bibitem[\protect\citeauthoryear{{Jaffe} et~al.,}{{Jaffe}
  et~al.}{2004}]{Jaf2004}
{Jaffe} W.,  et~al., 2004, \mn@doi [Nature] {10.1038/nature02531}, \href
  {http://adsabs.harvard.edu/abs/2004Natur.429...47J} {429, 47}

\bibitem[\protect\citeauthoryear{{Jansen}, {Fabricant}, {Franx}  \&
  {Caldwell}}{{Jansen} et~al.}{2000}]{Jan2000}
{Jansen} R.~A.,  {Fabricant} D.,  {Franx} M.,   {Caldwell} N.,  2000, \mn@doi
  [\apjs] {10.1086/313308}, \href
  {https://ui.adsabs.harvard.edu/abs/2000ApJS..126..331J} {126, 331}

\bibitem[\protect\citeauthoryear{{Jones} et~al.,}{{Jones}
  et~al.}{2009}]{Jon2009}
{Jones} D.~H.,  et~al., 2009, \mn@doi [Monthly Notices of the Royal
  Astronomical Society] {10.1111/j.1365-2966.2009.15338.x}, \href
  {http://adsabs.harvard.edu/abs/2009MNRAS.399..683J} {399, 683}

\bibitem[\protect\citeauthoryear{{Kelly}, {Sobolewska}  \&
  {Siemiginowska}}{{Kelly} et~al.}{2011}]{Kel2011}
{Kelly} B.~C.,  {Sobolewska} M.,   {Siemiginowska} A.,  2011, in American
  Astronomical Society Meeting Abstracts \#217. p. 142.33

\bibitem[\protect\citeauthoryear{{Kennicutt}}{{Kennicutt}}{1992}]{Ken1992}
{Kennicutt} Robert~C. J.,  1992, \mn@doi [\apjs] {10.1086/191653}, \href
  {https://ui.adsabs.harvard.edu/abs/1992ApJS...79..255K} {79, 255}

\bibitem[\protect\citeauthoryear{{Kennicutt} \& {Keel}}{{Kennicutt} \&
  {Keel}}{1984}]{Ken1984}
{Kennicutt} R.~C. J.,  {Keel} W.~C.,  1984, \mn@doi [\apjl] {10.1086/184243},
  \href {https://ui.adsabs.harvard.edu/abs/1984ApJ...279L...5K} {279, L5}

\bibitem[\protect\citeauthoryear{{Kewley}, {Heisler}, {Dopita}  \&
  {Lumsden}}{{Kewley} et~al.}{2001}]{Kew20012}
{Kewley} L.~J.,  {Heisler} C.~A.,  {Dopita} M.~A.,   {Lumsden} S.,  2001,
  \mn@doi [\apjs] {10.1086/318944}, \href
  {https://ui.adsabs.harvard.edu/abs/2001ApJS..132...37K} {132, 37}

\bibitem[\protect\citeauthoryear{{Khachikian} \& {Weedman}}{{Khachikian} \&
  {Weedman}}{1971}]{Kha1971}
{Khachikian} E.~Y.,  {Weedman} D.~W.,  1971, \mn@doi [\apjl] {10.1086/180701},
  \href
  {https://ui-adsabs-harvard-edu.ezproxy.lib.monash.edu.au/abs/1971ApJ...164L.109K}
  {164, L109}

\bibitem[\protect\citeauthoryear{{Khachikian}, {Asatrian}  \&
  {Burenkov}}{{Khachikian} et~al.}{2011}]{Kha2011}
{Khachikian} E.~Y.,  {Asatrian} N.~S.,   {Burenkov} A.~N.,  2011, \mn@doi
  [Astrophysics] {10.1007/s10511-011-9155-z}, \href
  {https://ui-adsabs-harvard-edu.ezproxy.lib.monash.edu.au/abs/2011Ap.....54...26K}
  {54, 26}

\bibitem[\protect\citeauthoryear{{Kim}, {Sanders}, {Veilleux}, {Mazzarella}  \&
  {Soifer}}{{Kim} et~al.}{1995}]{Kim1995}
{Kim} D.~C.,  {Sanders} D.~B.,  {Veilleux} S.,  {Mazzarella} J.~M.,   {Soifer}
  B.~T.,  1995, \mn@doi [\apjs] {10.1086/192157}, \href
  {https://ui.adsabs.harvard.edu/abs/1995ApJS...98..129K} {98, 129}

\bibitem[\protect\citeauthoryear{{Kokubo}}{{Kokubo}}{2015}]{Kok2015}
{Kokubo} M.,  2015, \mn@doi [Monthly Notices of the Royal Astronomical Society]
  {10.1093/mnras/stv241}, \href
  {http://adsabs.harvard.edu/abs/2015MNRAS.449...94K} {449, 94}

\bibitem[\protect\citeauthoryear{{Kollatschny} \& {Fricke}}{{Kollatschny} \&
  {Fricke}}{1987}]{Kol1987}
{Kollatschny} W.,  {Fricke} K.~J.,  1987, \aap, \href
  {https://ui.adsabs.harvard.edu/abs/1987A&A...183....9K} {183, 9}

\bibitem[\protect\citeauthoryear{{Koss} et~al.,}{{Koss} et~al.}{2017}]{Kos2017}
{Koss} M.,  et~al., 2017, \mn@doi [\apj] {10.3847/1538-4357/aa8ec9}, \href
  {https://ui.adsabs.harvard.edu/abs/2017ApJ...850...74K} {850, 74}

\bibitem[\protect\citeauthoryear{{LaMassa} et~al.,}{{LaMassa}
  et~al.}{2015}]{Lam2015}
{LaMassa} S.~M.,  et~al., 2015, \mn@doi [Astrophysical Journal]
  {10.1088/0004-637X/800/2/144}, \href
  {http://adsabs.harvard.edu/abs/2015ApJ...800..144L} {800, 144}

\bibitem[\protect\citeauthoryear{{Lira}, {Johnson}, {Lawrence}  \& {Cid Fernand
  es}}{{Lira} et~al.}{2007}]{Lir2007}
{Lira} P.,  {Johnson} R.~A.,  {Lawrence} A.,   {Cid Fernand es} R.,  2007,
  \mn@doi [\mnras] {10.1111/j.1365-2966.2007.12006.x}, \href
  {https://ui.adsabs.harvard.edu/abs/2007MNRAS.382.1552L} {382, 1552}

\bibitem[\protect\citeauthoryear{{MacLeod}, {Ross}, {Lawrence}, {Goad},
  {Horne}, {Burgett}  \& {Chambers}}{{MacLeod} et~al.}{2016}]{Mac2016}
{MacLeod} C.~L.,  {Ross} N.~P.,  {Lawrence} A.,  {Goad} M.,  {Horne} K.,
  {Burgett} W.,   {Chambers} K.~C.,  2016, \mn@doi [Monthly Notices of the
  Royal Astronomical Society] {10.1093/mnras/stv2997}, \href
  {http://adsabs.harvard.edu/abs/2016MNRAS.457..389M} {457, 389}

\bibitem[\protect\citeauthoryear{{MacLeod} et~al.,}{{MacLeod}
  et~al.}{2019}]{Mac2019}
{MacLeod} C.~L.,  et~al., 2019, \mn@doi [\apj] {10.3847/1538-4357/ab05e2},
  \href
  {https://ui-adsabs-harvard-edu.ezproxy.lib.monash.edu.au/abs/2019ApJ...874....8M}
  {874, 8}

\bibitem[\protect\citeauthoryear{{Maia}, {da Costa}, {Willmer}, {Pellegrini}
  \& {Rite}}{{Maia} et~al.}{1987}]{Mai1987}
{Maia} M.~A.~G.,  {da Costa} L.~N.,  {Willmer} C.,  {Pellegrini} P.~S.,
  {Rite} C.,  1987, \mn@doi [\aj] {10.1086/114336}, \href
  {https://ui.adsabs.harvard.edu/abs/1987AJ.....93..546M} {93, 546}

\bibitem[\protect\citeauthoryear{{Maia}, {Suzuki}, {da Costa}, {Willmer}  \&
  {Rite}}{{Maia} et~al.}{1996}]{Mai1996}
{Maia} M.~A.~G.,  {Suzuki} J.~A.,  {da Costa} L.~N.,  {Willmer} C.~N.~A.,
  {Rite} C.,  1996, \aaps, \href
  {https://ui.adsabs.harvard.edu/abs/1996A&AS..117..487M} {117, 487}

\bibitem[\protect\citeauthoryear{Mainzer et~al.,}{Mainzer
  et~al.}{2014}]{Mai2014}
Mainzer A.,  et~al., 2014, \mn@doi [The Astrophysical Journal]
  {10.1088/0004-637x/792/1/30}, 792, 30

\bibitem[\protect\citeauthoryear{{Marchese}, {Braito}, {Della Ceca},
  {Caccianiga}  \& {Severgnini}}{{Marchese} et~al.}{2012}]{Mar2012}
{Marchese} E.,  {Braito} V.,  {Della Ceca} R.,  {Caccianiga} A.,   {Severgnini}
  P.,  2012, \mn@doi [Monthly Notices of the Royal Astronomical Society]
  {10.1111/j.1365-2966.2012.20445.x}, \href
  {http://adsabs.harvard.edu/abs/2012MNRAS.421.1803M} {421, 1803}

\bibitem[\protect\citeauthoryear{{Marin}, {Porquet}, {Goosmann},
  {Dov{\v{c}}iak}, {Muleri}, {Grosso}  \& {Karas}}{{Marin}
  et~al.}{2013}]{Mar2013}
{Marin} F.,  {Porquet} D.,  {Goosmann} R.~W.,  {Dov{\v{c}}iak} M.,  {Muleri}
  F.,  {Grosso} N.,   {Karas} V.,  2013, \mn@doi [\mnras]
  {10.1093/mnras/stt1677}, \href
  {https://ui-adsabs-harvard-edu.ezproxy.lib.monash.edu.au/abs/2013MNRAS.436.1615M}
  {436, 1615}

\bibitem[\protect\citeauthoryear{{M{\'a}rquez} et~al.,}{{M{\'a}rquez}
  et~al.}{2004}]{Mar2004}
{M{\'a}rquez} I.,  et~al., 2004, \mn@doi [\aap] {10.1051/0004-6361:20034108},
  \href {https://ui.adsabs.harvard.edu/abs/2004A&A...416..475M} {416, 475}

\bibitem[\protect\citeauthoryear{{Masetti} et~al.,}{{Masetti}
  et~al.}{2006a}]{Mas2006}
{Masetti} N.,  et~al., 2006a, \mn@doi [\aap] {10.1051/0004-6361:20054332},
  \href {https://ui.adsabs.harvard.edu/abs/2006A&A...449.1139M} {449, 1139}

\bibitem[\protect\citeauthoryear{{Masetti} et~al.,}{{Masetti}
  et~al.}{2006b}]{Mas20062}
{Masetti} N.,  et~al., 2006b, \mn@doi [\aap] {10.1051/0004-6361:20066055},
  \href {https://ui.adsabs.harvard.edu/abs/2006A&A...459...21M} {459, 21}

\bibitem[\protect\citeauthoryear{{Meisenheimer} et~al.,}{{Meisenheimer}
  et~al.}{2007}]{Mei2007}
{Meisenheimer} K.,  et~al., 2007, \mn@doi [Astronomy & Astrophysics]
  {10.1051/0004-6361:20066967}, \href
  {http://adsabs.harvard.edu/abs/2007A%26A...471..453M} {471, 453}

\bibitem[\protect\citeauthoryear{{Merloni} et~al.,}{{Merloni}
  et~al.}{2015}]{Mer2015}
{Merloni} A.,  et~al., 2015, \mn@doi [\mnras] {10.1093/mnras/stv1095}, \href
  {https://ui-adsabs-harvard-edu.ezproxy.lib.monash.edu.au/abs/2015MNRAS.452...69M}
  {452, 69}

\bibitem[\protect\citeauthoryear{{Moran}, {Halpern}  \& {Helfand}}{{Moran}
  et~al.}{1994}]{Mor1994}
{Moran} E.~C.,  {Halpern} J.~P.,   {Helfand} D.~J.,  1994, \mn@doi [\apjl]
  {10.1086/187549}, \href
  {https://ui.adsabs.harvard.edu/abs/1994ApJ...433L..65M} {433, L65}

\bibitem[\protect\citeauthoryear{{Moran}, {Halpern}  \& {Helfand}}{{Moran}
  et~al.}{1996}]{Mor1996}
{Moran} E.~C.,  {Halpern} J.~P.,   {Helfand} D.~J.,  1996, \mn@doi
  [Astrophysical Journal, Supplement] {10.1086/192341}, \href
  {http://adsabs.harvard.edu/abs/1996ApJS..106..341M} {106, 341}

\bibitem[\protect\citeauthoryear{{Morris} \& {Ward}}{{Morris} \&
  {Ward}}{1988}]{Mor1988}
{Morris} S.~L.,  {Ward} M.~J.,  1988, \mn@doi [\mnras]
  {10.1093/mnras/230.4.639}, \href
  {https://ui.adsabs.harvard.edu/abs/1988MNRAS.230..639M} {230, 639}

\bibitem[\protect\citeauthoryear{{Moustakas} \& {Kennicutt}}{{Moustakas} \&
  {Kennicutt}}{2006}]{Mou2006}
{Moustakas} J.,  {Kennicutt} Jr. R.~C.,  2006, \mn@doi [\apjs]
  {10.1086/500971}, \href {http://adsabs.harvard.edu/abs/2006ApJS..164...81M}
  {164, 81}

\bibitem[\protect\citeauthoryear{{Nagayama}}{{Nagayama}}{2012}]{Nag2012}
{Nagayama} T.,  2012, African Skies, \href
  {https://ui-adsabs-harvard-edu.ezproxy.lib.monash.edu.au/abs/2012AfrSk..16...98N}
  {16, 98}

\bibitem[\protect\citeauthoryear{{Noda} \& {Done}}{{Noda} \&
  {Done}}{2018}]{Nod2018}
{Noda} H.,  {Done} C.,  2018, \mn@doi [Monthly Notices of the Royal
  Astronomical Society] {10.1093/mnras/sty2032}, \href
  {http://adsabs.harvard.edu/abs/2018MNRAS.tmp.1938N} {}

\bibitem[\protect\citeauthoryear{{Oknyansky} et~al.,}{{Oknyansky}
  et~al.}{2017}]{Okn2017}
{Oknyansky} V.~L.,  et~al., 2017, \mn@doi [Monthly Notices of the Royal
  Astronomical Society] {10.1093/mnras/stx149}, \href
  {http://adsabs.harvard.edu/abs/2017MNRAS.467.1496O} {467, 1496}

\bibitem[\protect\citeauthoryear{{Oknyansky}, {Lipunov}, {Gorbovskoy},
  {Winkler}, {van Wyk}, {Tsygankov}  \& {Buckley}}{{Oknyansky}
  et~al.}{2018}]{Okn2018}
{Oknyansky} V.~L.,  {Lipunov} V.~M.,  {Gorbovskoy} E.~S.,  {Winkler} H.,  {van
  Wyk} F.,  {Tsygankov} S.,   {Buckley} D.~A.~H.,  2018, The Astronomer's
  Telegram, \href {http://adsabs.harvard.edu/abs/2018ATel11915....1O} {11915}

\bibitem[\protect\citeauthoryear{{Osterbrock}}{{Osterbrock}}{1977}]{Ost1977}
{Osterbrock} D.~E.,  1977, \mn@doi [\apj] {10.1086/155407}, \href
  {https://ui.adsabs.harvard.edu/abs/1977ApJ...215..733O} {215, 733}

\bibitem[\protect\citeauthoryear{{Osterbrock}}{{Osterbrock}}{1981}]{Ost1981}
{Osterbrock} D.~E.,  1981, \mn@doi [The Astrophysics Journal] {10.1086/159306},
  \href {http://adsabs.harvard.edu/abs/1981ApJ...249..462O} {249, 462}

\bibitem[\protect\citeauthoryear{{Osterbrock}}{{Osterbrock}}{1985}]{Ost1985}
{Osterbrock} D.~E.,  1985, \mn@doi [\pasp] {10.1086/131487}, \href
  {https://ui.adsabs.harvard.edu/abs/1985PASP...97...25O} {97, 25}

\bibitem[\protect\citeauthoryear{{Osterbrock}}{{Osterbrock}}{1989}]{Ost1989}
{Osterbrock} D.~E.,  1989, {Astrophysics of gaseous nebulae and active galactic
  nuclei}.
University Science Books

\bibitem[\protect\citeauthoryear{{Owen}, {Ledlow}  \& {Keel}}{{Owen}
  et~al.}{1996}]{Owe1996}
{Owen} F.~N.,  {Ledlow} M.~J.,   {Keel} W.~C.,  1996, \mn@doi [\aj]
  {10.1086/117759}, \href
  {https://ui.adsabs.harvard.edu/abs/1996AJ....111...53O} {111, 53}

\bibitem[\protect\citeauthoryear{{Padovani} et~al.,}{{Padovani}
  et~al.}{2017}]{Pad2017}
{Padovani} P.,  et~al., 2017, \mn@doi [Astronomy & Astrophysics Reviews]
  {10.1007/s00159-017-0102-9}, \href
  {http://adsabs.harvard.edu/abs/2017A%26ARv..25....2P} {25, 2}

\bibitem[\protect\citeauthoryear{{Penston} \& {Perez}}{{Penston} \&
  {Perez}}{1984}]{Pen1984}
{Penston} M.~V.,  {Perez} E.,  1984, \mn@doi [Monthly Notices of the Royal
  Astronomical Society] {10.1093/mnras/211.1.33P}, \href
  {http://adsabs.harvard.edu/abs/1984MNRAS.211P..33P} {211, 33P}

\bibitem[\protect\citeauthoryear{{Phillips}, {Charles}  \&
  {Baldwin}}{{Phillips} et~al.}{1983}]{Phi1983}
{Phillips} M.~M.,  {Charles} P.~A.,   {Baldwin} J.~A.,  1983, \mn@doi [\apj]
  {10.1086/160797}, \href
  {https://ui.adsabs.harvard.edu/abs/1983ApJ...266..485P} {266, 485}

\bibitem[\protect\citeauthoryear{{Pietsch}, {Bischoff}, {Boller},
  {Doebereiner}, {Kollatschny}  \& {Zimmermann}}{{Pietsch}
  et~al.}{1998}]{Pie1998}
{Pietsch} W.,  {Bischoff} K.,  {Boller} T.,  {Doebereiner} S.,  {Kollatschny}
  W.,   {Zimmermann} H.~U.,  1998, \aap, \href
  {https://ui.adsabs.harvard.edu/abs/1998A&A...333...48P} {333, 48}

\bibitem[\protect\citeauthoryear{{Ramos Almeida}, {Mart{\'\i}nez Gonz{\'a}lez},
  {Asensio Ramos}, {Acosta-Pulido}, {H{\"o}nig}, {Alonso-Herrero}, {Tadhunter}
  \& {Gonz{\'a}lez-Mart{\'\i}n}}{{Ramos Almeida} et~al.}{2016}]{Ram2016}
{Ramos Almeida} C.,  {Mart{\'\i}nez Gonz{\'a}lez} M.~J.,  {Asensio Ramos} A.,
  {Acosta-Pulido} J.~A.,  {H{\"o}nig} S.~F.,  {Alonso-Herrero} A.,  {Tadhunter}
  C.~N.,   {Gonz{\'a}lez-Mart{\'\i}n} O.,  2016, \mn@doi [\mnras]
  {10.1093/mnras/stw1388}, \href
  {https://ui.adsabs.harvard.edu/abs/2016MNRAS.461.1387R} {461, 1387}

\bibitem[\protect\citeauthoryear{{Reichardt}, {Jimenez}  \&
  {Heavens}}{{Reichardt} et~al.}{2001}]{Rei2001}
{Reichardt} C.,  {Jimenez} R.,   {Heavens} A.~F.,  2001, \mn@doi [\mnras]
  {10.1046/j.1365-8711.2001.04768.x}, \href
  {https://ui.adsabs.harvard.edu/abs/2001MNRAS.327..849R} {327, 849}

\bibitem[\protect\citeauthoryear{{Reimers}, {Koehler}  \& {Wisotzki}}{{Reimers}
  et~al.}{1996}]{Rei1996}
{Reimers} D.,  {Koehler} T.,   {Wisotzki} L.,  1996, \aaps, \href
  {https://ui.adsabs.harvard.edu/abs/1996A&AS..115..235R} {115, 235}

\bibitem[\protect\citeauthoryear{{Reunanen}, {Kotilainen}  \&
  {Prieto}}{{Reunanen} et~al.}{2003}]{Reu2003}
{Reunanen} J.,  {Kotilainen} J.~K.,   {Prieto} M.~A.,  2003, \mn@doi [\mnras]
  {10.1046/j.1365-8711.2003.06771.x}, \href
  {https://ui.adsabs.harvard.edu/abs/2003MNRAS.343..192R} {343, 192}

\bibitem[\protect\citeauthoryear{Ross et~al.,}{Ross et~al.}{2018}]{Ros2018}
Ross N.~P.,  et~al., 2018, \mn@doi [Monthly Notices of the Royal Astronomical
  Society] {10.18524/1810-4215.2017.30.114366}

\bibitem[\protect\citeauthoryear{{Rossa}, {van der Marel}, {B{\"o}ker},
  {Gerssen}, {Ho}, {Rix}, {Shields}  \& {Walcher}}{{Rossa}
  et~al.}{2006}]{Ros2006}
{Rossa} J.,  {van der Marel} R.~P.,  {B{\"o}ker} T.,  {Gerssen} J.,  {Ho}
  L.~C.,  {Rix} H.-W.,  {Shields} J.~C.,   {Walcher} C.-J.,  2006, \mn@doi
  [\aj] {10.1086/505968}, \href
  {https://ui.adsabs.harvard.edu/abs/2006AJ....132.1074R} {132, 1074}

\bibitem[\protect\citeauthoryear{{Ruan} et~al.,}{{Ruan} et~al.}{2016}]{Rua2016}
{Ruan} J.~J.,  et~al., 2016, \mn@doi [\apj] {10.3847/0004-637X/826/2/188},
  \href
  {https://ui-adsabs-harvard-edu.ezproxy.lib.monash.edu.au/abs/2016ApJ...826..188R}
  {826, 188}

\bibitem[\protect\citeauthoryear{{Rudy}, {Cohen}  \& {Ake}}{{Rudy}
  et~al.}{1988}]{Rud1988}
{Rudy} R.~J.,  {Cohen} R.~D.,   {Ake} T.~B.,  1988, \mn@doi [\apj]
  {10.1086/166642}, \href
  {https://ui.adsabs.harvard.edu/abs/1988ApJ...332..172R} {332, 172}

\bibitem[\protect\citeauthoryear{{Runnoe} et~al.,}{{Runnoe}
  et~al.}{2016}]{Run2016}
{Runnoe} J.~C.,  et~al., 2016, \mn@doi [Monthly Notices of the Royal
  Astronomical Society] {10.1093/mnras/stv2385}, \href
  {http://adsabs.harvard.edu/abs/2016MNRAS.455.1691R} {455, 1691}

\bibitem[\protect\citeauthoryear{{Sabbadin}, {Cappellaro}, {Salvadori}  \&
  {Turatto}}{{Sabbadin} et~al.}{1989}]{Sab1989}
{Sabbadin} F.,  {Cappellaro} E.,  {Salvadori} L.,   {Turatto} M.,  1989,
  \mn@doi [\apjl] {10.1086/185593}, \href
  {https://ui.adsabs.harvard.edu/abs/1989ApJ...347L...5S} {347, L5}

\bibitem[\protect\citeauthoryear{{Scarpa}, {Falomo}  \& {Pesce}}{{Scarpa}
  et~al.}{1996}]{Sca1996}
{Scarpa} R.,  {Falomo} R.,   {Pesce} J.~E.,  1996, \aaps, \href
  {https://ui.adsabs.harvard.edu/abs/1996A&AS..116..295S} {116, 295}

\bibitem[\protect\citeauthoryear{{Scarsi}}{{Scarsi}}{1997}]{Sca1997}
{Scarsi} L.,  1997, in Data Analysis in Astronomy. pp 65--78

\bibitem[\protect\citeauthoryear{{Schmidt}, {Ferreiro}, {Vega Neme}  \&
  {Oio}}{{Schmidt} et~al.}{2016}]{Sch2016}
{Schmidt} E.~O.,  {Ferreiro} D.,  {Vega Neme} L.,   {Oio} G.~A.,  2016, \mn@doi
  [\aap] {10.1051/0004-6361/201629343}, \href
  {https://ui.adsabs.harvard.edu/abs/2016A&A...596A..95S} {596, A95}

\bibitem[\protect\citeauthoryear{Senarath, Brown, Cluver, Jarrett  \&
  Ross}{Senarath et~al.}{2019}]{me}
Senarath M.~R.,  Brown M. J.~I.,  Cluver M.~E.,  Jarrett T.~H.,   Ross N.~P.,
  2019, \mn@doi [Research Notes of the {AAS}] {10.3847/2515-5172/ab191d}, 3, 62

\bibitem[\protect\citeauthoryear{{Seyfert}}{{Seyfert}}{1943}]{Sey1943}
{Seyfert} C.~K.,  1943, \mn@doi [Astrophysical Journal] {10.1086/144488}, \href
  {http://adsabs.harvard.edu/abs/1943ApJ....97...28S} {97, 28}

\bibitem[\protect\citeauthoryear{{Shapovalova} et~al.,}{{Shapovalova}
  et~al.}{2019}]{Sha2019}
{Shapovalova} A.~I.,  et~al., 2019, \mn@doi [\mnras] {10.1093/mnras/stz692},
  \href
  {https://ui-adsabs-harvard-edu.ezproxy.lib.monash.edu.au/abs/2019MNRAS.485.4790S}
  {485, 4790}

\bibitem[\protect\citeauthoryear{{Shappee} et~al.,}{{Shappee}
  et~al.}{2014}]{Sha2014}
{Shappee} B.~J.,  et~al., 2014, \mn@doi [\apj] {10.1088/0004-637X/788/1/48},
  \href {http://adsabs.harvard.edu/abs/2014ApJ...788...48S} {788, 48}

\bibitem[\protect\citeauthoryear{{Siemiginowska}, {Czerny}  \&
  {Kostyunin}}{{Siemiginowska} et~al.}{1996}]{Sie1996}
{Siemiginowska} A.,  {Czerny} B.,   {Kostyunin} V.,  1996, \mn@doi [\apj]
  {10.1086/176831}, \href
  {https://ui.adsabs.harvard.edu/abs/1996ApJ...458..491S} {458, 491}

\bibitem[\protect\citeauthoryear{{Stepanian}, {Chavushyan}, {Carrasco},
  {Vald{\'e}s}, {M{\'u}jica}, {Tovmassian}  \& {Ayvazyan}}{{Stepanian}
  et~al.}{2002}]{Ste2002}
{Stepanian} J.~A.,  {Chavushyan} V.~H.,  {Carrasco} L.,  {Vald{\'e}s} J.~R.,
  {M{\'u}jica} R.~M.,  {Tovmassian} H.~M.,   {Ayvazyan} V.~T.,  2002, \mn@doi
  [\aj] {10.1086/341953}, \href
  {https://ui.adsabs.harvard.edu/abs/2002AJ....124.1283S} {124, 1283}

\bibitem[\protect\citeauthoryear{{Stern} et~al.,}{{Stern}
  et~al.}{2018}]{Ste2018}
{Stern} D.,  et~al., 2018, \mn@doi [\apj] {10.3847/1538-4357/aac726}, \href
  {http://adsabs.harvard.edu/abs/2018ApJ...864...27S} {864, 27}

\bibitem[\protect\citeauthoryear{{Stoklasov{\'a}}, {Ferruit}, {Emsellem},
  {Jungwiert}, {P{\'e}contal}  \& {S{\'a}nchez}}{{Stoklasov{\'a}}
  et~al.}{2009}]{Sto2009}
{Stoklasov{\'a}} I.,  {Ferruit} P.,  {Emsellem} E.,  {Jungwiert} B.,
  {P{\'e}contal} E.,   {S{\'a}nchez} S.~F.,  2009, \mn@doi [\aap]
  {10.1051/0004-6361/200811225}, \href
  {https://ui.adsabs.harvard.edu/abs/2009A&A...500.1287S} {500, 1287}

\bibitem[\protect\citeauthoryear{{Storchi Bergmann}, {Bica}  \&
  {Pastoriza}}{{Storchi Bergmann} et~al.}{1990}]{Sto1990}
{Storchi Bergmann} T.,  {Bica} E.,   {Pastoriza} M.~G.,  1990, \mnras, \href
  {https://ui.adsabs.harvard.edu/abs/1990MNRAS.245..749S} {245, 749}

\bibitem[\protect\citeauthoryear{{Storchi-Bergmann}, {Baldwin}  \&
  {Wilson}}{{Storchi-Bergmann} et~al.}{1993}]{Sto1993}
{Storchi-Bergmann} T.,  {Baldwin} J.~A.,   {Wilson} A.~S.,  1993, \mn@doi
  [\apjl] {10.1086/186867}, \href
  {https://ui.adsabs.harvard.edu/abs/1993ApJ...410L..11S} {410, L11}

\bibitem[\protect\citeauthoryear{{Thomas} et~al.,}{{Thomas}
  et~al.}{2017}]{Tho2017}
{Thomas} A.~D.,  et~al., 2017, \mn@doi [\apjs] {10.3847/1538-4365/aa855a},
  \href {https://ui.adsabs.harvard.edu/abs/2017ApJS..232...11T} {232, 11}

\bibitem[\protect\citeauthoryear{{Tohline} \& {Osterbrock}}{{Tohline} \&
  {Osterbrock}}{1976}]{Toh1976}
{Tohline} J.~E.,  {Osterbrock} D.~E.,  1976, \mn@doi [Astrophysical Journal,
  Letters] {10.1086/182317}, \href
  {http://adsabs.harvard.edu/abs/1976ApJ...210L.117T} {210, L117}

\bibitem[\protect\citeauthoryear{{Trakhtenbrot} et~al.,}{{Trakhtenbrot}
  et~al.}{2019}]{Tra2019}
{Trakhtenbrot} B.,  et~al., 2019, \mn@doi [\apj] {10.3847/1538-4357/ab39e4},
  \href {https://ui.adsabs.harvard.edu/abs/2019ApJ...883...94T} {883, 94}

\bibitem[\protect\citeauthoryear{{Tran}, {Osterbrock}  \& {Martel}}{{Tran}
  et~al.}{1992a}]{Tra1992}
{Tran} H.~D.,  {Osterbrock} D.~E.,   {Martel} A.,  1992a, \mn@doi [Astronomical
  Journal] {10.1086/116382}, \href
  {http://adsabs.harvard.edu/abs/1992AJ....104.2072T} {104, 2072}

\bibitem[\protect\citeauthoryear{{Tran}, {Miller}  \& {Kay}}{{Tran}
  et~al.}{1992b}]{Tra19922}
{Tran} H.~D.,  {Miller} J.~S.,   {Kay} L.~E.,  1992b, \mn@doi [\apj]
  {10.1086/171801}, \href
  {https://ui.adsabs.harvard.edu/abs/1992ApJ...397..452T} {397, 452}

\bibitem[\protect\citeauthoryear{Trippe, Crenshaw, Deo  \& Dietrich}{Trippe
  et~al.}{2008}]{Tri2008}
Trippe M.~L.,  Crenshaw D.~M.,  Deo R.,   Dietrich M.,  2008, \mn@doi [The
  Astronomical Journal] {10.1088/0004-6256/135/6/2048}, 135, 2048

\bibitem[\protect\citeauthoryear{Trippe, Crenshaw, Deo, Dietrich, Kraemer,
  Rafter  \& Turner}{Trippe et~al.}{2010}]{Tri2010}
Trippe M.~L.,  Crenshaw D.~M.,  Deo R.~P.,  Dietrich M.,  Kraemer S.~B.,
  Rafter S.~E.,   Turner T.~J.,  2010, \mn@doi [The Astrophysical Journal]
  {10.1088/0004-637x/725/2/1749}, 725, 1749

\bibitem[\protect\citeauthoryear{{Tsalmantza} et~al.,}{{Tsalmantza}
  et~al.}{2009}]{Tsa2009}
{Tsalmantza} P.,  et~al., 2009, \mn@doi [\aap] {10.1051/0004-6361/200912014},
  \href {https://ui.adsabs.harvard.edu/abs/2009A&A...504.1071T} {504, 1071}

\bibitem[\protect\citeauthoryear{{Veron-Cetty} \& {Veron}}{{Veron-Cetty} \&
  {Veron}}{1986a}]{Ver1986}
{Veron-Cetty} M.~P.,  {Veron} P.,  1986a, \aaps, \href
  {https://ui.adsabs.harvard.edu/abs/1986A&AS...65..241V} {65, 241}

\bibitem[\protect\citeauthoryear{{Veron-Cetty} \& {Veron}}{{Veron-Cetty} \&
  {Veron}}{1986b}]{Ver19862}
{Veron-Cetty} M.~P.,  {Veron} P.,  1986b, \aaps, \href
  {https://ui.adsabs.harvard.edu/abs/1986A&AS...66..335V} {66, 335}

\bibitem[\protect\citeauthoryear{{V{\'e}ron-Cetty} \&
  {V{\'e}ron}}{{V{\'e}ron-Cetty} \& {V{\'e}ron}}{2003}]{Ver2003}
{V{\'e}ron-Cetty} M.-P.,  {V{\'e}ron} P.,  2003, \mn@doi [\aap]
  {10.1051/0004-6361:20034225}, \href
  {https://ui.adsabs.harvard.edu/abs/2003A%26A...412..399V} {412, 399}

\bibitem[\protect\citeauthoryear{{V{\'e}ron-Cetty} \&
  {V{\'e}ron}}{{V{\'e}ron-Cetty} \& {V{\'e}ron}}{2010}]{Ver2010}
{V{\'e}ron-Cetty} M.-P.,  {V{\'e}ron} P.,  2010, \mn@doi [\aap]
  {10.1051/0004-6361/201014188}, \href
  {http://adsabs.harvard.edu/abs/2010A\%26A...518A..10V} {518, A10}

\bibitem[\protect\citeauthoryear{{Weedman}}{{Weedman}}{1976}]{Wee1976}
{Weedman} D.~W.,  1976, \qjras, \href
  {https://ui.adsabs.harvard.edu/abs/1976QJRAS..17..227W} {17, 227}

\bibitem[\protect\citeauthoryear{{Wei}, {Xu}, {Dong}  \& {Hu}}{{Wei}
  et~al.}{1999}]{Wei1999}
{Wei} J.~Y.,  {Xu} D.~W.,  {Dong} X.~Y.,   {Hu} J.~Y.,  1999, \mn@doi [\aaps]
  {10.1051/aas:1999514}, \href
  {https://ui.adsabs.harvard.edu/abs/1999A&AS..139..575W} {139, 575}

\bibitem[\protect\citeauthoryear{{White} et~al.,}{{White}
  et~al.}{2000}]{Whi2000}
{White} R.~L.,  et~al., 2000, \mn@doi [\apjs] {10.1086/313300}, \href
  {https://ui.adsabs.harvard.edu/abs/2000ApJS..126..133W} {126, 133}

\bibitem[\protect\citeauthoryear{{Winkler}}{{Winkler}}{1992}]{Win1992}
{Winkler} H.,  1992, \mn@doi [\mnras] {10.1093/mnras/257.4.677}, \href
  {https://ui.adsabs.harvard.edu/abs/1992MNRAS.257..677W} {257, 677}

\bibitem[\protect\citeauthoryear{{Wolf} et~al.,}{{Wolf} et~al.}{2018}]{Wol2018}
{Wolf} C.,  et~al., 2018, \mn@doi [\pasa] {10.1017/pasa.2018.5}, \href
  {http://adsabs.harvard.edu/abs/2018PASA...35...10W} {35, e010}

\bibitem[\protect\citeauthoryear{{Wright} et~al.,}{{Wright}
  et~al.}{2010}]{Wri2010}
{Wright} E.~L.,  et~al., 2010, \mn@doi [\aj] {10.1088/0004-6256/140/6/1868},
  \href {https://ui.adsabs.harvard.edu/abs/2010AJ....140.1868W} {140, 1868}

\bibitem[\protect\citeauthoryear{{Yang} et~al.,}{{Yang} et~al.}{2018}]{Yan2018}
{Yang} Q.,  et~al., 2018, \mn@doi [\apj] {10.3847/1538-4357/aaca3a}, \href
  {https://ui-adsabs-harvard-edu.ezproxy.lib.monash.edu.au/abs/2018ApJ...862..109Y}
  {862, 109}

\bibitem[\protect\citeauthoryear{{Zetzl} et~al.,}{{Zetzl}
  et~al.}{2018}]{Zel2018}
{Zetzl} M.,  et~al., 2018, \mn@doi [\aap] {10.1051/0004-6361/201732506}, \href
  {https://ui.adsabs.harvard.edu/abs/2018A&A...618A..83Z} {618, A83}

\bibitem[\protect\citeauthoryear{{de Grijp}, {Keel}, {Miley}, {Goudfrooij}  \&
  {Lub}}{{de Grijp} et~al.}{1992}]{Gri1992}
{de Grijp} M.~H.~K.,  {Keel} W.~C.,  {Miley} G.~K.,  {Goudfrooij} P.,   {Lub}
  J.,  1992, \aaps, \href
  {https://ui.adsabs.harvard.edu/abs/1992A&AS...96..389D} {96, 389}

\bibitem[\protect\citeauthoryear{{de Ruiter} \& {Lub}}{{de Ruiter} \&
  {Lub}}{1986}]{Rui1986}
{de Ruiter} H.~R.,  {Lub} J.,  1986, in {Swarup} G.,  {Kapahi} V.~K.,  eds,
  IAU Symposium Vol. 119, Quasars. p.~89

\makeatother
\end{thebibliography}







\bsp	
\label{lastpage}
\end{document}